\documentclass[journal,12pt,onecolumn,draftclsnofoot,]{IEEEtran}
\IEEEoverridecommandlockouts
\usepackage{cite}
\usepackage{amsmath,amssymb,amsfonts}
\usepackage{algorithmic}
\usepackage{graphicx}
\usepackage{diagbox}
\usepackage{textcomp}
\usepackage{xcolor}
\usepackage{scalerel}
\usepackage{csquotes}
\usepackage{dsfont}
\usepackage{subcaption}
\usepackage{float}
\usepackage{amsmath,amsthm}
\usepackage[]{graphicx}
\DeclareMathOperator*{\argmax}{argmax}
\usepackage{lipsum}
\def\BibTeX{{\rm B\kern-.05em{\sc i\kern-.025em b}\kern-.08em
    T\kern-.1667em\lower.7ex\hbox{E}\kern-.125emX}}

\begin{document}

\title{Information-Centric Grant-Free Access for IoT Fog Networks: Edge vs Cloud Detection and Learning\\
\thanks{This work has received funding from the European  Research  Council (ERC) under the European Union Horizon 2020 research and innovation program (grant agreements 725731 and 648382).}
}

\author{\IEEEauthorblockN{Rahif Kassab\IEEEauthorrefmark{1},
Osvaldo Simeone\IEEEauthorrefmark{1} and Petar Popovski\IEEEauthorrefmark{2} }\\
\IEEEauthorblockA{\small\IEEEauthorrefmark{1}Centre for Telecommunications Research, King's College London, London, United Kingdom\\
\IEEEauthorrefmark{2}Department of Electronic Systems, Aalborg University, Aalborg, Denmark\\
Emails: \IEEEauthorrefmark{1}\{rahif.kassab,osvaldo.simeone\}@kcl.ac.uk,
\IEEEauthorrefmark{2}petarp@es.aau.dk}}

\maketitle

\begin{abstract}
A multi-cell Fog-Radio Access Network (F-RAN) architecture is considered in which Internet of Things (IoT) devices periodically make noisy observations of a Quantity of Interest (QoI) and transmit using grant-free access in the uplink. The devices in each cell are connected to an Edge Node (EN), which may also have a finite-capacity fronthaul link to a central processor. In contrast to conventional information-agnostic protocols, the devices transmit using a Type-Based Multiple Access (TBMA) protocol that is tailored to enable the estimate of the field of correlated QoIs in each cell based on the measurements received from IoT devices. In this paper, this form of information-centric radio access is studied for the first time in a multi-cell F-RAN model with edge or cloud
detection. Edge and cloud detection are designed and compared for a multi-cell system. Optimal model-based detectors are introduced and the resulting asymptotic behavior of the probability of error at cloud and edge is derived. Then, for the scenario in which a statistical model is not available, data-driven edge and cloud detectors are discussed and evaluated in numerical results.
\end{abstract}

\begin{IEEEkeywords}
5G, IoT, Grant-Free Access, Type-Based Multiple Access, Fog-RAN, Machine-Type Communications, Information-Centric Access
\end{IEEEkeywords}

\section{Introduction}
\label{sec:introduction}
\subsection{Context}
Most commercial Internet of Things (IoT) systems are currently based on proprietary \textcolor{black}{protocols}, most notably LoRa \cite{lora} and Sigfox \cite{sigfox}\cite{IoT_B_Lorenzo}, and target long-range low-duty cycle transmission \cite{5g2016view}\cite{popovski2018slicing}. With the advent of 5G, cellular systems are expected to play an increasing role in IoT systems, thanks to the introduction of NarrowBand IoT (NB-IoT) \cite{NB_IoT_1}. IoT deployments based on cellular systems come with potential advantages in terms of reliability and coverage, but they also pose a number of novel challenges, particularly in terms of interference management and system optimization. \par A key communication primitive for IoT systems is grant-free access, whereby devices transmit using randomly selected preambles \cite{grant_free}\cite{gianluigi2015csda}. Random access is agnostic to the information being communicated, since all packets are generally treated in the same way as independent messages. {\color{black}{In this paper, we observe that preambles in IoT systems can be repurposed to serve as building blocks for a Type-Based Multiple Access (TBMA) protocol enabling remote estimation \cite{mergen2006tbma,anandkumar2007type,tbma_sayeed}. We use this observation to introduce an information-centric protocol based on TBMA that obtains a highly efficient grant-free access scheme.}}

\begin{figure*}
\centering
\begin{subfigure}{.5\textwidth}
  \centering
  \includegraphics[height= 5 cm, width= 7 cm]{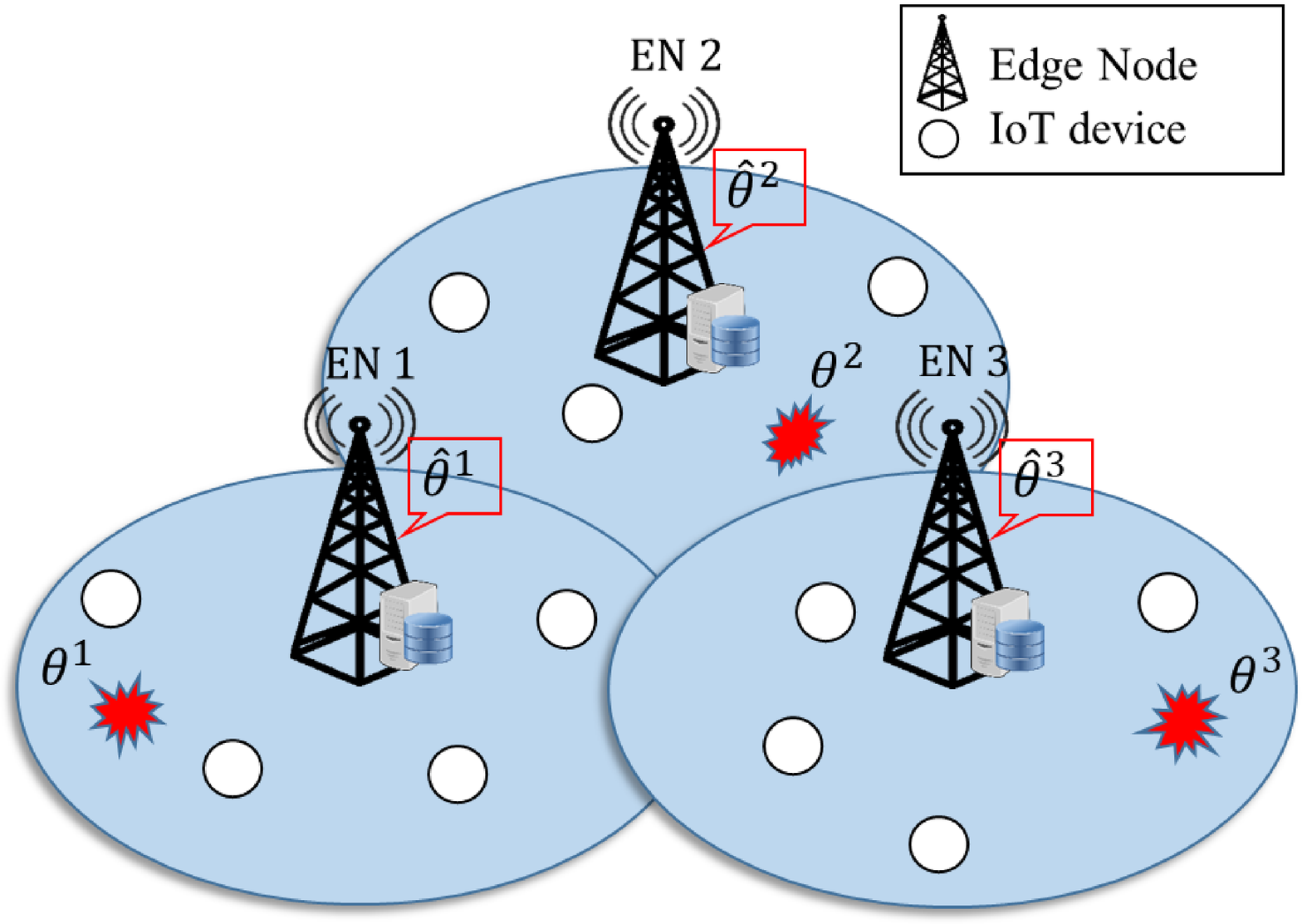}
  \caption{Edge detection}
  \label{fig:edge_detection}
\end{subfigure}%
\begin{subfigure}{.5\textwidth}
  \centering
  \includegraphics[height= 7 cm, width= 7 cm]{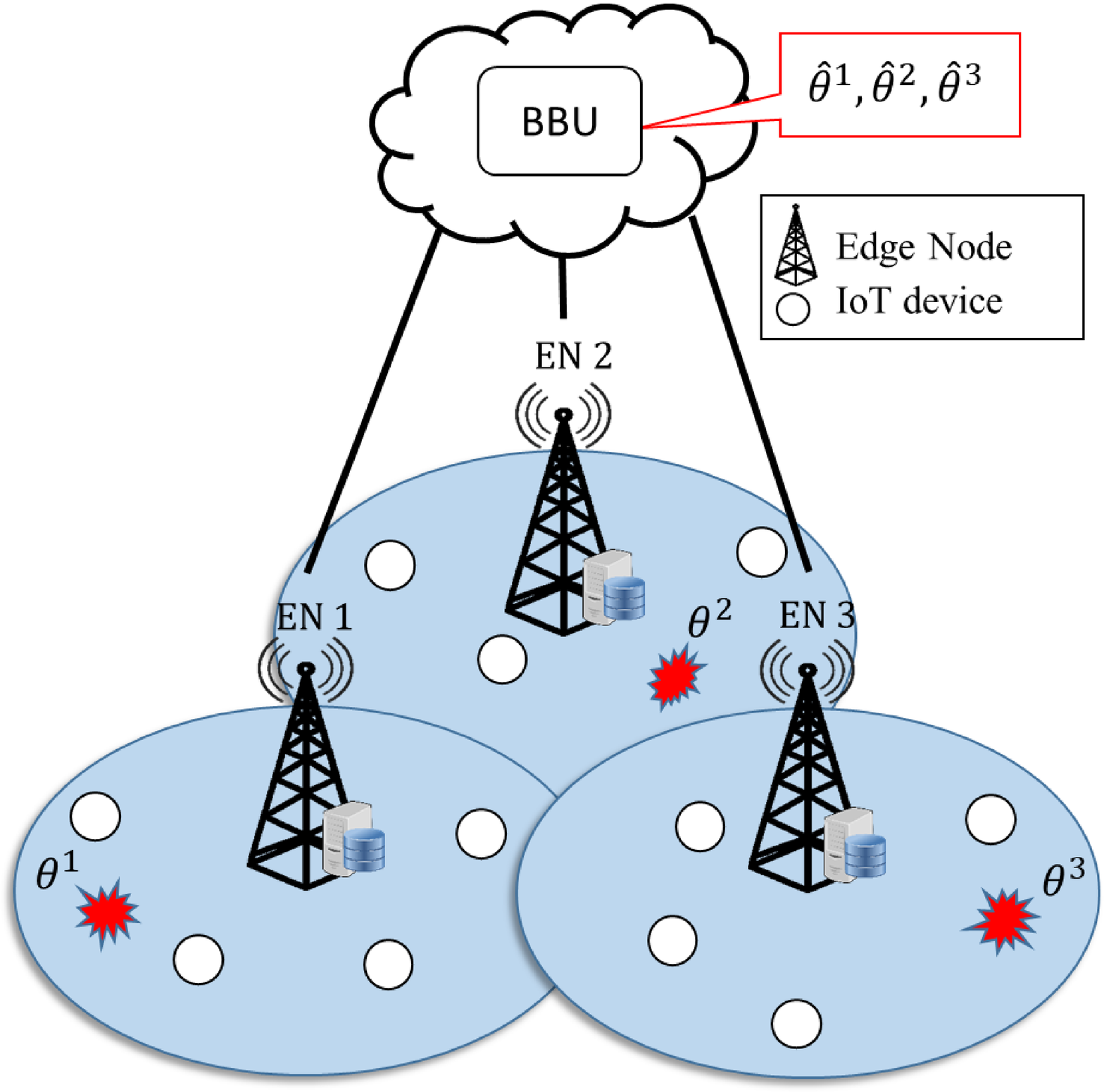}
  \caption{Cloud detection}
  \label{fig:cloud_detection}
\end{subfigure}
\caption{A multi-cell fog radio access network with IoT devices making observations of local quantities of interest (QoIs) $\theta^c$ for each cell $c$. Each cell uses the same frequency band. The goal of the system is to compute an estimate $\hat{\theta}^c$ for each $\theta^c$. This can be done in: (a) a distributed fashion at each EN, or (b) a centralized fashion at the cloud.}
\label{fig:system_model_multi_cell}
\end{figure*}

To define the problem of interest, as illustrated in Fig.~\ref{fig:system_model_multi_cell}, \textcolor{black}{we} consider an IoT application that aims at detecting the spatial distribution, of field, defined by a given Quantity of Interest (QoI) $\theta^c$ in each cell $c$. As an example, the IoT network may be deployed to monitor the pollution level across the covered geographical area. IoT devices operate as sensors that observe generally correlated information given that QoIs measured in nearby locations are likely to be similar. A conventional approach, implemented for instance in Sigfox, is to have each device transmit its observation using grant-free access to the local Edge Node (EN), which estimates the given QoI based on the received observations. This solution has a number of drawbacks that we address in this paper, namely: 
\begin{itemize}
    \item The communication protocol does not account for the correlation in the devices' observations and for the fact that the goal of the system is not to retrieve individual observations, but rather to estimate the field of QoIs;
    \item Local detection at the EN does not leverage the possible availability of central, or ``cloud'', processors that are connected to multiple ENs via fronthaul links. The presence of cloud processors, also known as Central Units in 3GPP documents \cite{3gpp_ran}, define cellular architectures referred to here as Fog-Radio Access Network (F-RAN) as in, e.g., \cite{tandon2016harnessing}\cite{peng2016fog}.
\end{itemize}
\subsection{TBMA in F-RAN Systems}
With regards to the first point raised above, in this work we adopt an \textit{information-centric} TBMA-based protocol. TBMA is a random access technique introduced in \cite{mergen2006tbma} and \cite{tbma_sayeed} and further studied, among other papers, in \cite{anandkumar2007type}. TBMA relies on the fact that, in order to optimally estimate a given parameter, only the histogram of the parameter-dependent measurements is needed and not the individual observations of the devices. Therefore, conventional transmission schemes that aim at ensuring recovery of all individual observations at the receiver are, generally, inefficient. In contrast, TBMA is designed to allow the receiver to estimate the histogram of the observations across the devices. To this end, in TBMA, all devices that make the same measurement, upon suitable quantization \cite{tbma_sayeed}, transmit the same waveform in a non-orthogonal fashion to the receiver. Assigning orthogonal waveforms for each measurement value hence yields bandwidth requirements that do not scale with the number of devices but only with the size of the quantized observation space. This produces potentially dramatic savings in terms of bandwidth and overall power, particularly in the regime of large number of devices \cite{mergen2006tbma}\cite{tbma_sayeed}\cite{anandkumar2007type}. All prior work on TBMA assumed a single-cell scenario with a single receiver.\par
Concerning the second point, with 5G, the cellular architecture is evolving from a base station-centric architecture, which is characterized by local processing, to a fog-like set-up, in which network functionalities can be distributed more flexibly between centralized processing at the cloud and local processing at the edge. Enabling this flexibility are fronthaul links connecting ENs to the cloud processor and network softwarization. At one extreme of the resulting F-RAN architecture, all processing can be local, e.g., carried out at the ENs, while, at the other, all processing can be centralized as in a Cloud-Radio Access Network (C-RAN) \cite{bookcransimeone}\cite{simeone2012cooperative}. In an IoT network, it is hence interesting to investigate under which conditions a centralized, cloud-based, detection of the QoIs can be advantageous. The problem is non-trivial due to the limitations on the capacity of the fronthaul links (see, e.g., \cite{bookcransimeone} \cite{simeone2012cooperative}).\par
In this paper, as illustrated in Fig.~\ref{fig:system_model_multi_cell}, we investigate an information-centric TBMA-based access scheme for F-RAN IoT systems that integrates in-cell TBMA with inter-cell non-orthogonal frequency reuse in the presence of either edge or cloud detection.
\subsection{Related Work}
IoT systems have been studied \textcolor{black}{from a number of viewpoints, reflecting the variety of their use cases and deployments. A long line of work is concerned with understanding and designing random access schemes that aim at recovering either the individual messages sent by active devices and/or their identities. These schemes can typically leverage sparsity in the devices' activation \cite{compressed_sensing_1,compressed_sensing_2,compressed_sensing_3,compressed_sensing_4}, which are generally assumed to be uncorrelated. Studies range from information-theoretical analyses of unsourced random access \cite{unsourced_polyanskiy} to applications of machine learning \cite{destounis2019learn2mac,cohen_RL_DSA
,jiang2019cooperative}. All these works implicitly disregard any correlation in the devices' messages and adopt conventional separate source-channel coding techniques. Correlation among devices' message was recently considered in \cite{petar_massive_common} via a simple correlation model where all devices can observe a common alarm message.}\par
\textcolor{black}{The problem of distributed detection based on local observations to a fusion center has been widely studied in the literature on wireless sensor networks, which typically assumes orthogonal transmissions \cite{fusion_decisions_rayleigh_fading
, lai2010fusion, performance_analysis_fusion, diffusion_LMS }. As some illustrative examples, references  \cite{decision_fusion_MIMO_WSN
,perf_analysis_energy_detection_MIMO,rician_MIMO_jamming, optimality_received_energy_decision_fusion} considered the distributed detection problem in the presence of multiple antennas at the receiver, while cooperative transmission was studied as an alternative solution in \cite{ALJARRAH2017127}\cite{efficient_fusion_cooperative_networks}.}\par
TBMA can be interpreted as carrying out a special form of Non-Orthogonal Multiple Access (NOMA) in that the devices transmit using non-orthogonal waveforms. In this sense, it is also related to the unsourced model of random access studied in \cite{yuri_unsourced}. Unlike conventional NOMA (see, e.g., \cite{noma_saito,nomadaimagazine,ding2014performance}), in TBMA, the communication protocol is tailored to the information being transmitted and to the detection task. It can hence be interpreted as an example of \textit{joint source-channel coding}, which is more generally receiving renewed interest for its potential spectral and power efficiency in IoT systems (see, e.g., \cite{morteza_joint,deniz2019deep_source_channel,popovski2019start}). \textcolor{black}{A recent related work is \cite{JSC_MP_grant_free_IoT} that introduces a novel Bayesian Message Passing technique with joint source-channel coding via a non-orthogonal generalization of TBMA; while in \cite{liu_decision_hybrid_MAC} a hybrid orthogonal and non-orthogonal multiple access channel based on TBMA was introduced with an optimized decision rule}. Based on this review, to the best of our knowledge, TBMA has not been studied in multi-cell F-RAN systems.
\par The problem of studying the performance trade-offs between processing at the edge and at the cloud has been studied in a number of works, including for content delivery \cite{sengupta_content_delivery}\cite{zhang2019fundamental}, scheduling \cite{kang2018control}, and coexistence of different 5G services \cite{rahif_access_2018}.
\subsection{Main Contributions}
The main contributions of this paper are summarized as follows:
\begin{itemize}
    \item An information-centric grant free access scheme is introduced for F-RAN IoT cellular systems that combines in-cell TBMA and inter-cell non-orthogonal frequency reuse;
    \item Optimal edge and cloud detectors are derived for the system at hand that leverage correlations in the QoIs across different cells;
    \item An analytical study of the performance of optimal cloud and edge detection is provided in terms of detection error exponents;
    \item Assuming absence of model knowledge at the edge or cloud, learning-based data-driven detection schemes are considered for both cloud and edge processing.
\end{itemize}
The rest of the paper is organized as follows. In Sec. \ref{sec:system_signal_model} we detail both the system and the signal models. In Sec. \ref{sec:protocol_metrics} we highlight the communication protocol used by the devices in addition to the performance metrics utilized to evaluate the performance of the system. In Sec. \ref{sec:optimal_detection} and \ref{sec:asymptotic_performance} we study and analyze edge and cloud detection with optimal detection and the corresponding asymptotic behaviour respectively. In Sec. \ref{sec:machine_learning}, we investigate data-driven edge and cloud detection for the case where a statistical model is not available. Numerical results are presented in Sec. \ref{sec:numerical_results} and conclusions and extensions are proposed in Sec. \ref{sec:generalization}.
\par
\textbf{Notation:} Lower-case bold characters represent vectors and upper-case bold characters represent matrices. $\mathbf{A}^{\mathsf{T}}$ denotes the transpose of matrix $\mathbf{A}$. $|\mathbf{A}|$ denotes the determinant of matrix $\mathbf{A}$. $A(i,j)$ denotes the element of $\mathbf{A}$ located at the $i$-th row and $j$-th column. $\mathcal{CN}(x|\mu,\sigma^2)$ is the probability density function (pdf) of a complex Gaussian random variable (RV) with mean $\mu$ and standard deviation $\sigma$. $\mathcal{P}(x|\lambda)$ represents the probability mass function (pmf) of a Poisson RV with mean $\lambda$. $C(f_1||f_2)$ and $D(f_1 || f_2)$ represent the Chernoff information and the Kullback-Leibler (KL) divergence respectively for the probability distributions $f_1$ and $f_2$. Given $a < b$, $[a,b]$ represents the segment of values between $a$ and $b$.
\section{System and Signal model}
\label{sec:system_signal_model}
\subsection{System Model} 
As illustrated in Fig.~\ref{fig:system_model_multi_cell}, we study a multi-cell wireless fog network that aims at detecting a field of Quantities of Interest (QoIs), such as temperature or pollution level, based on signals received from IoT devices. Each cell contains a single-antenna Edge Node (EN) and multiple IoT devices. We assume that the QoI is described in each cell $c$ by a Random Variable (RV) $\theta^c$. RVs $\{\theta^c\}$ are generally correlated across cells, and each device in cell $c$ makes a noisy measurement of $\theta^c$. For example, QoI $\theta^c$ may represent the pollution level in the area covered by cell $c$. In this paper, we assume for simplicity of notation and analysis that each QoI can take two possible values $\theta_0$ and $\theta_1$. Continuing the example above, $\theta^c$ may represent a low ($\theta_0$) or high ($\theta_1$) pollution level in cell $c$.
Extensions to more general QoIs follow directly but at the cost of a more cumbersome notation and analysis as further discussed in Sec. \ref{sec:generalization}.
\par The IoT devices are interrogated periodically by their local EN over a number $L$ of collection intervals, which are synchronized across all cells.
In each collection interval, a number of devices in each cell $c$ transmit their measurements in the uplink using a grant-free access protocol based on Type-Based Multiple Access (TBMA) \cite{tbma_sayeed}\cite{anandkumar2007type}. Note that the random activation pattern assumed here can also model aspects such as discontinuous access to the QoI or to sufficient energy-communication resources at the devices. Mathematically, in any collection interval $l=1,\ldots,L$, each IoT device in cell $c$ is active probabilistically, {\color{black}{independently of the observation being sensed}}, so that the total number $N_l^c$ of devices active in collection interval $l$ in cell $c$ is a Poisson RV with mean $\lambda$ and probability mass function $\mathrm{Pr}[N^c_l = n]=\mathcal{P}(n|\lambda)$.
When active, a device transmits a noisy measurement of the local QoI $\theta^c$ in the uplink. All devices share the same spectrum and hence their transmissions generally interfere, both within the same cell and across different cells.
\par
We compare two different architectures for detection of the QoIs: 
\textit{(i) Edge detection:} Detection of each QoI $\theta^c$ is done locally at the EN in cell $c$ based on the uplink signals received from the IoT devices, producing a local estimate $\hat{\theta}^c$ (see Fig.~\ref{fig:edge_detection}); and \textit{(ii) Cloud detection:} The ENs are connected with orthogonal finite-capacity digital fronthaul links to a cloud processor with fronthaul capacity of $C\ \mathrm{[bit/s/Hz]}$. As in a C-RAN architecture \cite{simeone2012cooperative}, each EN forwards the received signal upon quantization to the cloud processor using the fronthaul link. Unlike conventional C-RAN systems, here the goal is for the cloud to compute estimates $\{ \hat{\theta}^c \}$ of all QoIs $\{ \theta^c \}$ (see Fig.~\ref{fig:cloud_detection}).
\par
\subsection{Signal Model}
When active, an IoT device $i$ in cell $c$ during the $l$-th collection observes a measurement $X_{i,l}^{c}$. We assume that the measurement takes values in an alphabet $\{1,2, \ldots, M \}$ of size $M$. If the observation is analog, measurement $X^c_{i,l}$ can be obtained upon quantization to $M$ levels. The problem of designing the quantizer is an interesting direction for future research (see Sec. \ref{sec:generalization}).
\textcolor{black}{For the purpose of this analysis, a number of levels $M$ may be translated into a mean-squared error due to quantization that scales as $2^{-M}$ using standard quantization where the mean-squared error is equal to $\Delta^2/12$ \cite{uniform_quantizers} with $\Delta$ being the step size of the uniform quantizer.}

The distribution of each observation $X_{i,l}^{c}$ depends on the underlying QoI as
\begin{equation}
\begin{aligned}
&\mathrm{Pr}[X^c_{i,l}=m|\theta^c=\theta_0]=p_0^c(m)\\
\mathrm{and }\ \ &\mathrm{Pr}[X^c_{i,l}=m|\theta^c=\theta_1]=p_1^c(m), \label{eq:dist}
\end{aligned}
\end{equation}
for $m=1,\ldots, M$.
In words, devices in cell $c$ make generally noisy measurements with $\theta^c$-dependent distributions $p^c_0(\cdot)$ and $p^c_1(\cdot)$. When conditioned on QoIs  $\{\theta^c\}$, measurements $X_{i,l}^{c}$ are i.i.d. across all values of the cell index $c$, device index $i$, and the collection index $l$.
\begin{figure}[t]
	\centering
	\includegraphics[height= 8 cm, width= 9 cm]{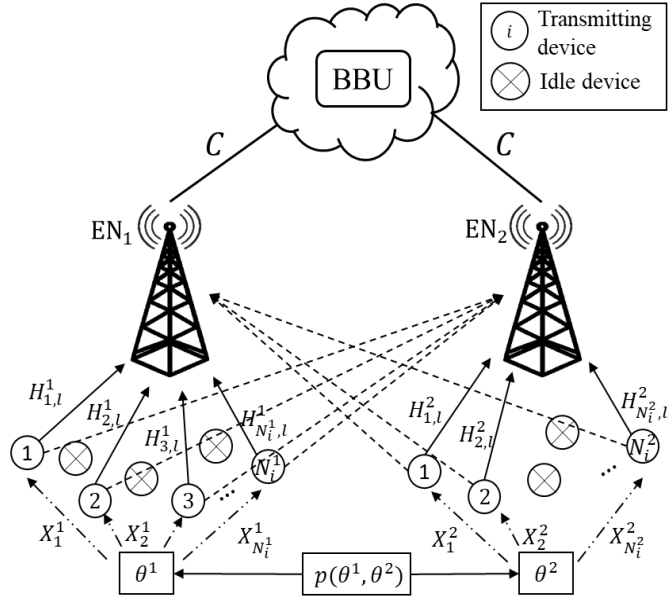}
	\caption{Two-cell system model. Dashed lines represent interference channels.}
	\label{fig:system_model}
\end{figure}
\par While the analysis can be generalized for a multi-cell scenario as further discussed in Sec. \ref{sec:generalization}, we henceforth focus on the two-cell case illustrated in Fig.~\ref{fig:system_model} in order to concentrate on the essence of the problem without complicating the notation. In this case, {\color{black}{we define the joint distribution of the QoIs in the two cells as}}
\textcolor{black}{
\begin{equation}
  p (\theta^1 , \theta^2) = \frac{\rho}{2} 1_{\{ \theta^1 = \theta^2\}} + \frac{1-\rho}{2} 1_{\{\theta^1 \neq \theta^2\}}   \label{eq:param_dist}
\end{equation}}
where $ 0 \leq \rho \leq 1$ represents a ``correlation" parameter that measures the probability that the two QoIs have the same value, i.e., $\rho = \mathrm{Pr} [\theta^1 = \theta^2]$. \textcolor{black}{In practice, the value of $\rho$ depends on the QoI and the size of the cells. However, this value is not needed at the receiver neither for detection nor decoding.}
Note that under \eqref{eq:param_dist}, both values of the QoI are equiprobable, i.e., $\mathrm{Pr}(\theta^c = \theta_j)=0.5$ for $j \in \{0,1\}$ and $c \in \{0,1\}$. \textcolor{black}{Furthermore, when $\rho=0.5$, the two QoIs are independent (these could correspond, e.g., to heat and pollution levels).} Extensions to more general probability distributions are immediate.
\par We denote by $H_{i,l}^c \sim \cal{CN}$ $(\mu_{H}, \sigma^{2}_{H})$ the flat-fading Ricean fading channel, with mean $\mu_H$ and variance $\sigma^2_H$, from device $i$ to the $\mathrm{EN}$ in the same cell $c$ during collection interval $l$; and by $G_{i,l}^{c} \sim \mathcal{CN}(\mu_{G}, \sigma^{2}_{G})$, with mean $\mu_G$ and variance $\sigma^2_G$, the flat-fading Ricean fading channel from device $i$ in cell $c^\prime \neq c$ to the EN in cell $c$ during collection interval $l$. All channels are assumed i.i.d. across indices $i,l$ and $c$. In the next section, we detail the communication protocol, including the physical-layer model and the performance metrics used.
\section{Communication protocol and Performance metrics}
\label{sec:protocol_metrics}
In this section, we detail the communication protocol and the performance metrics used to evaluate the system's performance.
\subsection{Communication Protocol}
As mentioned in Sec.~\ref{sec:introduction}, based on the single-cell results in \cite{mergen2006tbma,anandkumar2007type,tbma_sayeed}, in this paper we focus on an information-centric TBMA-based protocol that leverages the correlation between observations of different devices in different cells. To this end, within the available bandwidth and time per-collection interval, as in \cite{mergen2006tbma}, we assume the presence of $M$ orthogonal waveforms $\{ \phi_m(t), m= 1,\ldots,M\}$ with unit energy.
In practice, preambles allocated for the random access phase in cellular standards can be used as waveforms. These waveforms are used in a non-orthogonal fashion by the IoT devices to transmit their observations in the uplink. As detailed next, we allow for non-orthogonal frequency reuse across the two cells, and study also the orthogonal frequency reuse for comparison.
\par
\textit{Non-orthogonal frequency reuse:} According to TBMA, each waveform $\phi_m(t)$ encodes the value $m \in \{1,\ldots, M \}$ of the observations of a device. The signal transmitted by a device $i$ in cell $c$ that is active in interval $l$ is then given as
\begin{equation}
   S_{i,l}^c(t) = \sqrt{E_s} \phi_{X_{i,l}^c}(t), 
\end{equation}
that is, we have $S_{i,l}^c(t) = \sqrt{E_s} \phi_{m}(t)$ if the observed signal is $X_{i,l}^c(t)=m$, where $E_s$ is the transmission energy of a device per collection interval. With TBMA, devices observing the same value $m$ hence transmit using the same waveform. This is why, as discussed in Sec.~\ref{sec:introduction}, the spectral resources required by TBMA scale with the number $M$ of observations values rather than with the total amount of information by all the active devices, which may be much larger than $M$. \par 
The received signal at the $\mathrm{EN}$ in cell $c$ during the $l$-th collection can be written as
\begin{equation}
    Y_l^c(t) = \sum_{i=1}^{N_l^c} H_{i,l}^{c} S_{i,l}^c (t)+ \sum_{i=1}^{N_l^{c^\prime}} G_{i,l}^{c} S_{i,l}^{c^\prime}(t) + W_l^c(t), \label{eq:ylc}
\end{equation}
where $W_l(t)\sim\mathcal{CN}(0,W_0)$ is white Gaussian noise, i.i.d. over $l$ and $c$, with power $W_0$; and $c^\prime \neq c$ represents the index of the other cell. The first term in \eqref{eq:ylc} represents the contribution from the IoT devices in the same cell $c$, while the second term represents the contribution from IoT devices from the other cell $c^\prime$. \par
Given the orthogonality of the waveforms $\{ \phi_m (t) \}$, a demodulator based on a bank of matched filters can be implemented at each EN without loss of optimality \cite{anandkumar2007type}. After matched filtering of the received signal with all waveforms $\phi_{m}(t)$ for $m=1,\ldots,M$, each EN $c$ obtains the $M \times 1$ vector
\begin{equation}
\begin{aligned}
      \mathbf{Y}_l^c &= \frac{1}{\sqrt{E_s}} [\langle \phi_1(t),Y_l^c(t)\rangle, \ldots , \langle \phi_M(t),Y_l^c(t)\rangle]^{\mathsf{T}} \\ 
      &= \sum_{i=1}^{N_l^c} H_{i,l}^{c} \mathbf{e}_{X_{i,l}^c}+ \sum_{i=1}^{N_l^{c^\prime}} G_{i,l}^{c} \mathbf{e}_{X_{i,l}^{c^\prime}} + \mathbf{W}_l, \label{eq:Yc}\\
\end{aligned}
\end{equation}
where $\mathbf{W}_l$ is a vector with i.i.d. $\mathcal{CN}(0, \mathrm{SNR}^{-1})$ elements, with $\mathrm{SNR}= E_s/W_0$; and $\mathbf{e}_m$ represents an $M \times 1$ unit vector with all zero entries except in position $m$. In \eqref{eq:Yc}, we used the notation $\langle a(t),b(t)\rangle = \int a(t)b(t) dt$ to represent the correlation integral as applied to the given correlation interval.
To gain insight into the operation of TBMA, we note that, in the absence of noise and inter-cell interference, and if the channel coefficients are all equal one, i.e., with $\mu_G=\sigma^2_G = \sigma^2_H=0$ and $\mu_H=1$, the $m$-th element of vector $\mathbf{Y}^c_l$ is equal to the number of active devices that have observed the $m$-th data level in cell $c$ \cite{mergen2006tbma}.
\par
\textit{Orthogonal frequency reuse:} For reference, we also consider a rate-$1/2$ frequency reuse scheme that eliminates inter-cell interference. In this baseline scheme, the $M$  available orthogonal resources are equally partitioned between the two cells, so that in each cell only $M/2$ orthogonal waveforms are available. We assume here $M$ to be even for simplicity of notation. In this case, each active IoT device $i$ in cell $c$ quantizes its observation $X^c_{i,l}$ to $M/2$ levels as $\hat{X}^c_{i,l}=m$ if $X^c_{i,l} \in \{2m-1 , 2m \}$ for $m=1,\ldots,M/2$ before transmission.
The signal received at EN $c$ during collection $l$ can hence be written as 
\begin{equation}
\begin{aligned}
      \mathbf{Y}_l^c &= \frac{1}{\sqrt{E_s}} [\langle Y_l^c(t),\phi_1(t)\rangle, \ldots , \langle Y_l^c(t),\phi_{M/2}(t)\rangle]^{\mathsf{T}} \\ 
      &= \sum_{i=1}^{N_l^c} H_{i,l}^{c} \mathbf{e}_{\hat{X}_{i,l}^c} + \mathbf{W}_l^c. \label{eq:Yc_tdma}\\
\end{aligned}
\end{equation}
Comparing \eqref{eq:Yc_tdma} with \eqref{eq:Yc}, we observe that, on the one hand, orthogonal frequency reuse reduces the resolution of the observations of each device from $M$ to $M/2$ levels, but, on the other hand, it removes inter-cell interference. In the remainder of this paper, we consider and derive the performance of the more general non-orthogonal frequency reuse. The performance for orthogonal frequency reuse can be derived the same way by replacing the number of resources $M$ by $M/2$ and setting the interference channel coefficients to zero in all the derived equations. As for detection of the QoI, as illustrated in Fig.~\ref{fig:system_model_multi_cell}, we study both edge and cloud detection described as follows:

\textit{Edge Detection:} With edge detection, each EN $c$ produces an estimate $\hat{\theta^c}$ of the RV $\theta^c$ based on the received signals $\mathbf{Y}^c_l$ for all collection intervals $l=1,\ldots,L$, where $\mathbf{Y}^l_c$ is given in \eqref{eq:Yc} and \eqref{eq:Yc_tdma} for non-orthogonal and orthogonal frequency reuse, respectively. \par
\textit{Cloud Detection:} With cloud detection, each EN $c$ compresses the received signals $\{ \mathbf{Y}^c_l \}^{L}_{l=1}$ across all $L$ collection intervals and sends the resulting compressed signals $\{ \hat{\mathbf{Y}}^c_l \}_{l=1}^{L}$ to the cloud. Compression is needed in order to account for the finite fronthaul capacity $C$. The cloud carries out joint detection of both QoIs $\{\theta^1,\theta^2 \}$ producing estimates $\{ \Hat{\theta}^1 , \Hat{\theta}^2 \}$. 
\subsection{Performance Metrics} The performance of cloud and edge detection methods will be evaluated in terms of the joint error probability
\begin{equation}
\mathrm{P}_e = \mathrm{Pr}[\cup_{c=1}^{2} \{\hat{\theta^c} \neq \theta^c \}], \label{eq:joint_probability}    
\end{equation} 
where $\hat{\theta}^c$ is the estimate of the QoI $\theta^c$ obtained at $\mathrm{EN}$ $c$ or at the cloud, for edge detection and cloud detection respectively.
In order to enable analysis, we will also study analytically the scaling of the error probability $\mathrm{P}_e$ as a function of the number $L$ of collections. From large deviation theory, the detection error probability $P_e$ decays exponentially as \cite{cover2012elements}
\begin{equation}
    \mathrm{P}_e = \mathrm{exp}(-L E+ o(L))\ \ \ \mathrm{with}\ L\to\infty, \label{eq:Pe}
\end{equation}
where $o(L)/L \to 0$ as $L \to \infty$, for some detection error exponent $E$. We will hence be interested in computing analytically the error exponent $E$ for edge and cloud detection to verify our experimental results using optimal and machine learning based detection where $P_e$ is used as a performance metric. \textcolor{black}{In the latter, a finite number of collections $L$ is considered in order to capture realistic low-latency IoT scenarios.}\par
In the next two sections, we consider the case in which the model \eqref{eq:dist}-\eqref{eq:ylc} is available for the design of optimal detection at edge and cloud, and describe the resulting detectors and their asymptotic behavior in terms of the error probability via the error exponent when $L \to \infty$. Then, in Sec.~\ref{sec:machine_learning}, we study the case in which the detectors need to be learned from data rather than being derived from a mathematical model.
\section{Optimal Detection}
\label{sec:optimal_detection}
In this section, we assume that the joint distribution \eqref{eq:dist}-\eqref{eq:ylc} of the QoI, of the observations, and of the received signal is known, and we detail the corresponding optimal detectors at edge and cloud. The performance of these detectors is evaluated numerically in terms of the probability of error $P_e$ \eqref{eq:joint_probability} in Sec. \ref{sec:numerical_results}.
\subsection{Optimal Edge Detection}
\label{sec:optimal_edge}
With edge detection, each $\mathrm{EN}$ in cell $c$ performs the binary test
\begin{equation}
    \mathcal{H}_0^c: \theta^c = \theta_0\   \mathrm{versus}\   \mathcal{H}_1^c:\theta^c = \theta_1 \label{eq:test_1}
\end{equation}
based on the available received signals $\mathbf{Y}^c=\{\mathbf{Y}_l^c \}_{l=1}^{L}$ in \eqref{eq:Yc}.
The optimum Bayesian decision rule that minimizes the probability of error at each EN chooses the hypothesis with the Maximum A Posteriori (MAP) probability. Since the hypotheses in \eqref{eq:test_1} are a priori equiprobable the MAP rule is given by the log-likelihood ratio test:
\begin{equation}
       \log \frac{f(\mathbf{Y}^{c}| \theta^c = \theta_0)}{f(\mathbf{Y}^{c} | \theta^c = \theta_1)} \underset{{\hat{\theta}^c = \theta_1}}{\overset{\hat{\theta}^c = \theta_0}{\gtrless}} 0. \label{eq:optimal_edge}
\end{equation}
Using the law of total probability and the i.i.d. property across collection intervals $l$, the likelihood can be expressed as
\begin{equation}
\begin{aligned}
    f(\mathbf{Y}^c| \theta^c = \theta_j) = \sum_{k=0}^{1} \prod_{l=1}^L  & f(\mathbf{Y}^c_l | \theta^c = \theta_j, \theta^{c^\prime} = \theta_k) \\ &\times \mathrm{Pr}(\theta^{c^\prime} = \theta_k |\theta^c = \theta_j ), \label{eq:edge_dist}
    \end{aligned}
\end{equation}
where $\mathrm{Pr}(\theta^{c^\prime} = \theta_k |\theta^c = \theta_j) = 2\mathrm{Pr}(\theta^{c^\prime} = \theta_k,\theta^c = \theta_j)$ is the conditional probability of the QoI in cell $c^\prime$ obtained from \eqref{eq:param_dist}, and $f(\mathbf{Y}^c_l | \theta^c = \theta_j, \theta^{c^\prime} = \theta_k)$ represents the distribution of the signal \eqref{eq:ylc} received at EN $c$ during interval $l$ when we have $\theta^c = \theta_j$ and $\theta^{c^\prime}=\theta_k$. This distribution can be written as
\begin{equation}
\begin{aligned}
    f(\mathbf{Y}^c_l | \theta^c = \theta_j,&  \theta^{c^{\prime}} = \theta_k) 
    = \prod_{m=1}^M\sum_{n_1 = 0}^{\infty} \sum_{n_2 = 0}^{\infty} \mathcal{P}(n_1 | \lambda p_j^c (m)) \\ & \times \mathcal{P}(n_2 | \lambda p_k^{c^\prime} (m)) \mathcal{CN}(Y^c_l (m) | \mu_{n_1,n_2}, \sigma^2_{n_1,n_2}), \label{eq:y_dist_edge}
    \end{aligned}
\end{equation}
where we have defined
\begin{equation}
\mu_{n_1,n_2} = n_1 \mu_H + n_2 \mu_G,\ \ \mathrm{and} \ \sigma^2_{n_1,n_2}=n_1 \sigma_H^2 + n_2 \sigma_G^2  + W_0.
\end{equation}
The distribution \eqref{eq:y_dist_edge} follows since: \textit{(i)} conditioned on the numbers $n_1$ and $n_2$ of active devices in cell $c$ and $c^\prime$, respectively, the distribution of $Y^c_l(t)$ in \eqref{eq:ylc} is complex Gaussian with mean $\mu_{n_1, n_2}$ and variance $\sigma^2_{n_1 , n_2}$; and \textit{(ii)} by the Poisson thinning property \cite{billingsley2008probability}, the average number of devices transmitting signal level $m$ in cell $c$ under hypothesis $\theta^c = \theta_j$ is equal to $\lambda p_j^c (m)$. 
\subsection{Optimal Cloud Detection}
\label{sec:optimal_cloud}
The cloud tackles the quaternary hypothesis testing problem of distinguishing among hypotheses $\mathcal{H}_{jk}:(\theta^1,\theta^2) = (\theta_j,\theta_k)$ for $j,k \in \{0,1 \}$ on the basis of the quantized signals $\{ \hat{\mathbf{Y}}_l \}^{L}_{l=1}$ received from both ENs on the fronthaul links.
The optimal test for deciding among multiple hypotheses is the Bayes MAP rule that chooses the hypothesis $\mathcal{H}_{j k}$ by solving the problem
\begin{equation}
        \argmax_{\{j,k\} \in \{ 0,1\}^2 }\Bigg\{ \log p(\theta_j, \theta_k) +  \sum_{l=1}^{L} \log f(\hat{\mathbf{Y}}_{l} | \theta^1=\theta_j,\theta^{2}=\theta_k) \Bigg\}, \label{eq:optimal_cloud}
\end{equation}
where the first term represents the prior probability of hypothesis $\mathcal{H}_{jk}$ while the second term represents the distribution of the compressed signals $\hat{\mathbf{Y}}_l = [(\hat{\mathbf{Y}}^1_l)^{\mathsf{T}},(\hat{\mathbf{Y}}^2_l)^{\mathsf{T}}]^{\mathsf{T}}$ sent on the fronthaul links. This is derived next.
\par Following a by now standard approach, see, e.g., \cite{bookcransimeone}\cite{parkandsimeone}, the impact of fronthaul quantization is modeled as an additional quantization noise. In particular, the signal received at the cloud from EN $c$ can be written accordingly as 
\begin{equation}
    \hat{\mathbf{Y}}_{l}^c = \mathbf{Y}^{c}_{l} + \mathbf{Q}^c_l,
\end{equation}
where $\mathbf{Q}^c_l$ represents the quantization noise vector. As in most prior references (see, e.g., \cite{bookcransimeone}\cite{parkandsimeone}), the quantization noise vector $\mathbf{Q}^c_l$ is assumed to have i.i.d. elements being normally distributed with zero mean and variance $\sigma^2_{q^c}$.

\textcolor{black}{This assumption is justified by the fact that a high-dimensional dithered lattice quantizer, such as Trellis Coded Quantization, preceded by a linear transform can obtain a Gaussian quantization noise with any desired quantization spectrum \cite{lattice_quantization_noise}
\cite{trellis_quantization}. Furthermore, reference \cite{validity_gaussian} demonstrates that the assumption of additive Gaussian quantization noise is also valid for uniform scalar quantizers when the input distribution is continuous. }\\
Furthermore, from rate-distortion theory, the fronthaul capacity constraint implies the following inequality \cite{parkandsimeone}, for each EN $c$
\begin{equation}
    M C \geq I(\mathbf{Y}^c_{l} ; \hat{\mathbf{Y}}^c_{l}). \label{eq:capacity_contraint}
\end{equation}
This is because the number of bits available to transmit each measurement $\hat{\mathbf{Y}}^c_l$ is given by $C$ bits per symbol, or equivalently per orthogonal spectral resource, that is, $MC$ bits in total for all $M$ resources. From \eqref{eq:capacity_contraint}, one can in principle derive the quantization noise power $\sigma^2_{q^c}$.
\par Evaluating the mutual information in \eqref{eq:capacity_contraint} directly is, however, made difficult by the non-Gaussianity of the received signals $\mathbf{Y}_l^c $. To tackle this issue, we bound the mutual information term in \eqref{eq:capacity_contraint} using the property that the Gaussian distribution maximizes the differential entropy under covariance constraints \cite{cover2012elements}, obtaining the following result.\par
\textit{Lemma 1:} The quantization noise power can be upper bounded as $\sigma^2_{q^c} \leq \Bar{\sigma}^2_{q^c}$, where $\Bar{\sigma}^2_{q^c}$ is obtained by solving the non-linear equation
\begin{equation}
\begin{aligned}
    MC & = \frac{1}{2} \sum_{m=1}^{M} \log  \\  & \Bigg( \frac{ \sum_{j=0}^{1} \sum_{k=0}^{1} \mathrm{Pr}( \theta^1 = \theta^1_j,\theta^2 = \theta^2_k) \Sigma^{c}_{j,k} (m,m) + \sigma_{q^c}^2}{(\sigma_{q^c}^2)^M} \Bigg). \label{eq:tosolve_quantization}
    \end{aligned}
\end{equation}
where
\begin{equation}
    \Sigma^c_{j,k}(m,m)= \sigma^2_H \lambda p_j^c(m)  + \sigma^2_G  \lambda p_k^c(m) + \frac{1}{\mathrm{SNR}} \label{eq:BigSigma}
\end{equation}
are the diagonal elements of the covariance matrix $\mathbf{\Sigma}^c_{j,k}$ of $\mathbf{Y}_c^l$ when $\theta^c = \theta_j$ and $\theta^{c^{\prime}} = \theta_k$.
\par \textit{Proof:} See Appendix \ref{sec:appendix_1} for details. \qed \\
Using \textit{Lemma 1}, the distribution of the received signal $f(\hat{\mathbf{Y}}_l | \theta^1=\theta_j, \theta^2 = \theta_k)$ in \eqref{eq:optimal_cloud} can be evaluated as in \eqref{eq:y_dist_edge} but with a variance of $\sigma^2_{n_1,n_2} + \sigma^2_{q^c}$ in lieu of $\sigma^2_{n_1,n_2}$ for each cell $c$.
\section{Asymptotic Performance}
\label{sec:asymptotic_performance}
In this section, we derive the error exponent $E$ in \eqref{eq:Pe} for the optimal detectors discussed in Sec. \ref{sec:optimal_detection} when the number of collection intervals $L$ grows to infinity. In order to simplify the analysis, as in \cite{anandkumar2007type}, we will take the assumption of large average number of active devices, i.e., of large $\lambda$. This scenario is practically relevant for scenarios such as massive Machine Type Communication systems (mMTC), with large devices' density \cite{5g2016view}. In Sec.~\ref{sec:numerical_results}, we will further validate the approach by means of numerical results for smaller values of $L$ and $\lambda$.
\subsection{Edge Detection}
\label{sec:edge_detection_asymptotic}
The error exponent $E$ in \eqref{eq:Pe} using edge detection can be lower bounded as shown in the following proposition. \par
\textit{Proposition 1:} Under the optimal Bayesian detector \eqref{eq:optimal_edge}, the error exponent $E$ in \eqref{eq:Pe} in the large-$\lambda$ regime and for any $0 < \rho < 1$ is lower bounded as $E \geq E^{edge} = \mathrm{min}_{c \in \{1,2\}}E^c$, where
\begin{equation}
\begin{aligned}
    E^c  & = \min_{k \in \{ 0,1\}} \max_{\alpha \in [0,1]} \\ & \Bigg[\frac{1}{2} \sum_{m=1}^{M} \log\Big(\frac{\alpha \Sigma_{1,k}^c(m,m) + (1-\alpha)\Sigma_{2,k}^c(m,m)}{(\Sigma_{1,\textcolor{black}{k}}^{c}(m,m))^{\alpha} (\Sigma_{2,k}^{c}(m,m))^{\textcolor{black}{1-\alpha}}} \Big) \\& +\frac{\alpha(1- \alpha)}{2} \sum_{m=1}^{M} \frac{(\mu_{1,k}^c(m) - \mu_{2,k}^c (m))^2}{(\alpha \Sigma_{1,k}^c(m,m) + (1-\alpha) \Sigma_{2,k}^c (m,m) )} \Bigg] \label{eq:expo_edge}
\end{aligned}
\end{equation}
with 
\begin{equation}
   \mu_{j,k}^c(m) = \mu_H \lambda p_j^c(m) + \mu_G \lambda  p_k^{c^\prime}(m)
   \label{eq:dist_edge}
\end{equation}
and $\Sigma^c_{j,k}(m,m)$ given in \eqref{eq:BigSigma} for $j,k \in \{0,1\}$, $m\in [1,\ldots, M]$ and $c^\prime \neq c \in \{1,2\}$.
\par \textit{Proof:} In a manner similar to \cite[Theorem 3]{anandkumar2007type}, the proof of the above theorem relies on the Central Limit Theorem (CLT) with random number of summands \cite[p. 369]{cover2012elements} and on the error exponent for optimal binary Bayesian detection based on the Chernoff Information \cite{cover2012elements}. We refer to Appendix \ref{sec:appendix_a} for details. \qed
\par The term in \eqref{eq:expo_edge} being optimized over $k$ corresponds to the Chernoff information \cite[Chapter 11]{cover2012elements} for the binary test between the distributions of the received signal $\mathbf{Y}^c_l$ under hypotheses $\theta^c = \theta_0$ and $\theta^c = \theta_1$ when $\theta^{c^\prime}=\theta_k$. In fact, for large values of $\lambda$, when $\theta^c = \theta_j$ and $\theta^{c^\prime} = \theta_k$, the received signal $\mathbf{Y}^c_l$ in \eqref{eq:Yc} can be shown to be approximately distributed as $\mathcal{CN}(\boldsymbol{\mu}_{j,k}^c,\mathbf{\Sigma}_{j,k}^c)$, with mean vector $\boldsymbol{\mu}_{j,k}^c = [\mu_{j,k}^c(1) , \ldots , \mu_{j,k}^c(M) ]^{\mathsf{T}}$ and diagonal covariance matrix $\mathbf{\Sigma}_{j,k}^c$ with diagonal elements $\Sigma_{j,k}^c (m,m)$.
\subsection{Cloud Detection}
\label{sec:cloud_detection}
Here we analyze the performance of joint detection at the cloud described in \eqref{eq:optimal_cloud} in terms of the error exponent $E$.\par 
\textit{Proposition 2:} Under the optimal detector \eqref{eq:optimal_cloud}, the error exponent $E$ in \eqref{eq:Pe} in the large-$\lambda$ regime for cloud detection can be lower bounded as $E \geq E^{cloud} = \mathrm{min}_{\{j,k\} \in \{0,1 \}^2} E_{j,k}$, where
\begin{equation}
\begin{aligned}
    E_{j,k} &=\min_{ \{j^\prime,k^\prime\} \neq \{j,k\}} \max_{\alpha \in [0,1]} \Big[ \frac{1}{2} \mathrm{log} \frac{|\alpha \mathbf{\Sigma}_{j,k} + (1-\alpha)\mathbf{\Sigma}_{j^\prime ,k^\prime}|}{|\mathbf{\Sigma}_{j,k}|^\alpha|\mathbf{\Sigma}_{j^\prime ,k^\prime}|^{1-\alpha}} \\&+ \frac{\alpha (1-\alpha)}{2}(\boldsymbol{\mu}_{j,k} - \boldsymbol{\mu}_{j^\prime ,k^\prime})^{\mathsf{T}}(\alpha \mathbf{\Sigma}_{j,k} + (1-\alpha)\mathbf{\Sigma}_{j^\prime ,k^\prime})^{-1} \\ & \times (\boldsymbol{\mu}_{j,k} - \boldsymbol{\mu}_{j^\prime, k^\prime}) \Big], \label{eq:expo_cloud}
    \end{aligned}
\end{equation}
where the $2M \times 1$ vector $\boldsymbol{\mu}_{j,k}$ is defined as
\begin{equation}
\begin{aligned}
    & \mu_{j,k}(m) = \mu_{j,k}^1(m)\ \ \mathrm{for}\ m=1,\ldots,M\\
    & \mathrm{and}\ \ \ \ \mu_{j,k}(m) = \mu_{k,j}^2(m)\ \ \mathrm{for }\ m=M+1,\ldots,2M,\label{eq:param_cloud}  \\
\end{aligned}
\end{equation}
where $\mu_{j,k}^c(m)$ is defined in \eqref{eq:dist_edge}, and the $2M \times 2M$ covariance matrix $\mathbf{\Sigma}_{j,k}$ is given as
\begin{equation}
\begin{aligned}
    &\Sigma_{j,k} (m,m) = \Sigma_{j,k}^1 (m,m) + \sigma^2_{q^1} \ \ \mathrm{for }\ m=1,\ldots,M,\\
    &\Sigma_{j,k} (m,m) = \Sigma_{k,j}^2 (m,m) + \sigma^2_{q^2} \ \ \mathrm{for }\ m=M+1,\ldots,2M,\\
    & \Sigma_{j,k} (m,M+m) = \Sigma_{j,k} (M+m,m) = \\& p_{j}^{1}(m)(1-p_{j}^{1}(m)) \lambda \mu_H \mu_G + p_{k}^{2}(m)(1 - p_{k}^{2}(m))\lambda \mu_H \mu_G\ \ \\ & \mathrm{for }\ m=1,\ldots,M, \label{eq:param_cloud_2}
    \end{aligned}
\end{equation}
where $\Sigma^c_{j,k}(m,m)$ is defined in \eqref{eq:BigSigma} and all other entries of matrix $\mathbf{\Sigma}_{j,k}$ are zero.\par
\textit{Proof:} The proof follows in a manner similar to \textit{Proposition 1} as we detail in Appendix \ref{sec:appendix_b}. \qed 
\par 
The term in \eqref{eq:expo_cloud} being optimized over $\{ j^\prime , k^{\prime}\}$ corresponds to the Chernoff information for the binary test between the distribution of the signal received at the cloud under hypotheses $(\theta^c = \theta_j,\theta^{c^\prime} = \theta_k)$ and $(\theta^c = \theta_{j^\prime},\theta^{c^\prime} = \theta_{k^\prime})$. As for edge detection, the signal received at the cloud under hypothesis $\mathcal{H}_{jk}$ is approximately distributed as $\mathcal{CN}(\boldsymbol{\mu}_{j,k},\mathbf{\Sigma}_{j,k})$, where the elements of the mean vector $\boldsymbol{\mu}_{j,k}$ and covariance matrix $\mathbf{\Sigma}_{j,k}$ are described in \eqref{eq:param_cloud} and \eqref{eq:param_cloud_2}. Note that, by \eqref{eq:param_cloud_2}, the signals received from cell $c$ and $c^\prime$ are correlated, when conditioned on any hypothesis $\mathcal{H}_{j,k}$, if channels have non-zero mean.
\textcolor{black}{
\subsection{Edge vs Cloud Detection}
\label{sec:edge_vs_cloud}
In this section, we prove that the performance of cloud detection is superior to edge detection in terms of error exponent as long as the inter-cell channel power gain power $\sigma^2_G$ is sufficiently large. The main result can be summarized in the following theorem.}
\par \textcolor{black}{\textit{Theorem 1: } The error exponents derived in \textit{Proposition 1} and \textit{Proposition 2} satisfy the following limits
\begin{equation}
    \lim_{\sigma^2_G \to \infty} E^{edge} = 0\ \ \mathrm{and}\ \ \lim_{\sigma^2_G \to \infty} E^{cloud} > 0. \label{eq:theorem_1}
\end{equation}}
\par \textcolor{black}{\textit{Proof: }The proof can be found in Appendix D.}\qed \par
\textcolor{black}{\textit{Theorem 1} implies that, for high inter-cell power gains, edge detection leads to vanishing small error exponent, while this is not the case for cloud detection. This demonstrates that edge detection is inter-cell interference limited, while this is not the case for cloud detection. In practice, as shown via numerical results in Sec. \ref{sec:numerical_results}, fairly low interference power levels are sufficient for cloud detection to outperform edge detection.} \par
\textcolor{black}{In Sec. \ref{sec:numerical_results}, the comparison between the error exponents for edge detection and cloud detection is done in terms of the lower bounds via numerical simulations. Furthermore, it was found that these lower bounds provide good insights on the performance of the system in the non-asymptotic regime with finite $L$ where real optimal detection is used.}
\section{Edge and Cloud Learning}
\label{sec:machine_learning}
In the previous sections, we have assumed that ENs and the cloud are aware of the joint distribution \eqref{eq:dist}-\eqref{eq:ylc} of the QoIs, observations, and received signals. As a result, the conditional distributions $f(\mathbf{Y}^c | \theta^c)$ are known at each EN $c$ and the distributions $f(\hat{\mathbf{Y}}| \theta^1, \theta^2)$ are known at the cloud for all values of the QoIs. These distributions are needed in order to implement the optimal detectors \eqref{eq:optimal_edge} and \eqref{eq:optimal_cloud} at the edge and cloud respectively. In contrast, in this section, we assume lack of knowledge of the aforementioned distributions and use data-driven learning-based techniques at the edge and the cloud in order to train edge and cloud detectors. The performance of these detectors is evaluated using the probability of error $P_e$, and it is compared with the optimal detectors' performance, in Sec. \ref{sec:numerical_results}.
\subsection{Edge Learning}
In order to enable the training of a binary classifier at each EN $c$, we assume the availability of a labeled training set for supervised learning. This data set is defined by $N$ i.i.d. observations $\{(\mathbf{Y}^c(n),\theta^c(n))\}$ for $n=1,\ldots,N$, where $\mathbf{Y}^c(n) = [(\mathbf{Y}^c_1(n))^{\mathsf{T}},\ldots,(\mathbf{Y}^c_L(n))^{\mathsf{T}}]^{\mathsf{T}}$ is the $ML \times 1$ vector of observations at EN $c$, which is distributed according to the unknown conditional distribution $f(\mathbf{Y}^c(n)|\theta^c(n))$ and $\theta^c(n) \in \{\theta_0 , \theta_1 \}$ is the binary QoI. \textcolor{black}{This data set can be obtained offline during a calibration phase that uses either direct measurements or synthetically generated data from an emulator of the radio environment of interest \cite{zappone_wireless_design_deep_learning}.}
Any binary classifier can be trained based on this data set in order to generalize the mapping between input $\mathbf{Y}^c$ and output $\theta^c$ outside the training set. For illustration, we consider a feedforward neural network, which is described through the functional relations (see, e.g., \cite{simeone2018brief} \cite{goodfellow2016deep})
\begin{equation}
\begin{aligned}
    & \mathbf{h}^1 = h(\mathbf{W}^1 \tilde{\mathbf{Y}}^c (n))\\
    & \mathbf{h}^b = h(\mathbf{W}^b \mathbf{h}^{b-1})\ \mathrm{for}\ b=1,\ldots,B\\
    & \mathrm{Pr}(\theta^c=\theta_1)=\sigma(\mathbf{w}^{B+1} \mathbf{h}^{B}), \label{eq:edge_classifier}
\end{aligned}
\end{equation}
where $B$ is the number of hidden layers; $\mathbf{h}^b$ represents the vector of outputs of the $b$-th hidden layer with weight matrix $\mathbf{W}^b$ for $b=1,\ldots,B$; $\mathbf{w}^{B+1}$ is the vector of weights for the last layer; $h(\cdot)$ is a non-linear function, here taken to be hyperbolic tangent \cite{goodfellow2016deep}; $\sigma(x)=1/(1+e^{-x})$ is the sigmoid function; and we have $\tilde{\mathbf{Y}}^c = [1 , (\mathbf{Y}^c)^{\mathsf{T}}]^{\mathsf{T}}$ as the input of the neural network. The output of the neural network provides the probability that the QoI is equal $\theta_1$ for the given weights $\{ \{ \mathbf{W}_b \}_{b=1}^{B} ,\mathbf{w}^{B+1} \}$. The neural network is trained to minimize the cross-entropy loss via the backpropagation algorithm. Details of this standard procedure can be found, e.g., in \cite{simeone2018brief} \cite{goodfellow2016deep}.
\subsection{Cloud Learning}
Unlike the ENs, the cloud needs to train a multi-class classifier in order to distinguish among the four hypotheses $ \mathcal{H}_{jk}:\  (\theta^1 , \theta^2)  = (\theta_j , \theta_k )$ for $j,k \in \{0,1\}$. To enable supervised learning, we assume the availability of a labelled training set defined by $N$ i.i.d. observations $\{(\hat{\mathbf{Y}}(n),\theta^1(n),\theta^2(n))\}$ for $n=1,\ldots,N$, where $\hat{\mathbf{Y}}(n) = [(\hat{\mathbf{Y}}_1(n))^\mathsf{T},\ldots,(\hat{\mathbf{Y}}_L(n))^\mathsf{T}]^{\mathsf{T}}$ is the $2ML \times 1$ vector of observations at the cloud, which is distributed according to the unknown joint distribution $f(\hat{\mathbf{Y}}(n)|\theta^1(n),\theta^2(n))$ and $(\theta^1 , \theta^2)$ are the QoIs for the two cells. While any multi-class classifier can be used, here we consider a classifier based on a neural network as discussed above. Unlike the classifier in \eqref{eq:edge_classifier}, the cloud-based classifier contains four output neurons with each neuron representing the probability of one of the four hypotheses. The output layer is defined as in \eqref{eq:edge_classifier} but with a softmax non-linearity in lieu of the sigmoid \cite{simeone2018brief}\cite{goodfellow2016deep}. Training is carried out by optimizing the cross-entropy criterion.
\section{Numerical Results}
\label{sec:numerical_results}
In this section, we discuss the performance of edge and cloud-based detection and learning as a function of different system parameters, such as inter-cell interference strength and fronthaul capacity, through numerical examples. For the optimal detectors described in Sec.~\ref{sec:optimal_detection}, which require knowledge of the measurements and channel models, we consider both the analytical performance in terms of error exponent derived in Sec. \ref{sec:asymptotic_performance} and the performance in the regime with a finite number $L$ of observations evaluated via Monte Carlo simulations. For the learning-based solution, we evaluate the performance under the system model discussed in Sec.~\ref{sec:system_signal_model} in order to ensure a fair comparison with model-based solutions. 
\par The system contains two cells as illustrated in Fig.~\ref{fig:system_model}, and unless specified otherwise, we set the system parameters as follows: average number of active devices per cell $\lambda = 4$; average SNR equal to $\mathrm{SNR} = 3\ \mathrm{dB}$; direct channel parameters $\mu_H =1$ and $\sigma^2_H = 1$; inter-cell channel parameters $\mu_G =1$ and $\sigma^2_G=1$; correlation between the QoIs in the two cells $\rho=0.85$; and number of observations levels $M=4$. 
\textcolor{black}{
The assumption of equal statistics for direct and inter-cell channel parameters reflects an ultra-dense network deployment as considered in \cite[Sec. III]{SE_dense_MIMO} \cite[Sec. IV.B]{fundamental_characteristics_UDSN}.} Furthermore, the conditional distributions of the observations for both cells are given for QoI value $\theta_0$ as $p_0^1(1)=p_0^2(1)=0.4$, $p_0^1(2)=p_0^2(2)=0.3$, $p_0^1(3)=p_0^2(3)=0.2$ and $p_0^1(4)=p_0^2(4)=0.1$ and for QoI value $\theta_1$ $p_1^1(m)=p_1^2(m)=p_0^1(M-m+1)$. 
Note that, under QoI $\theta_0$, devices in both cells tend to measurements with small values $m$, while the opposite is true under QoI $\theta_1$. For example, value $\theta_0$ may represent a low pollution level or temperature.
\begin{figure}[h]
	\centering
	\includegraphics[height= 6.3 cm, width= 9.5 cm]{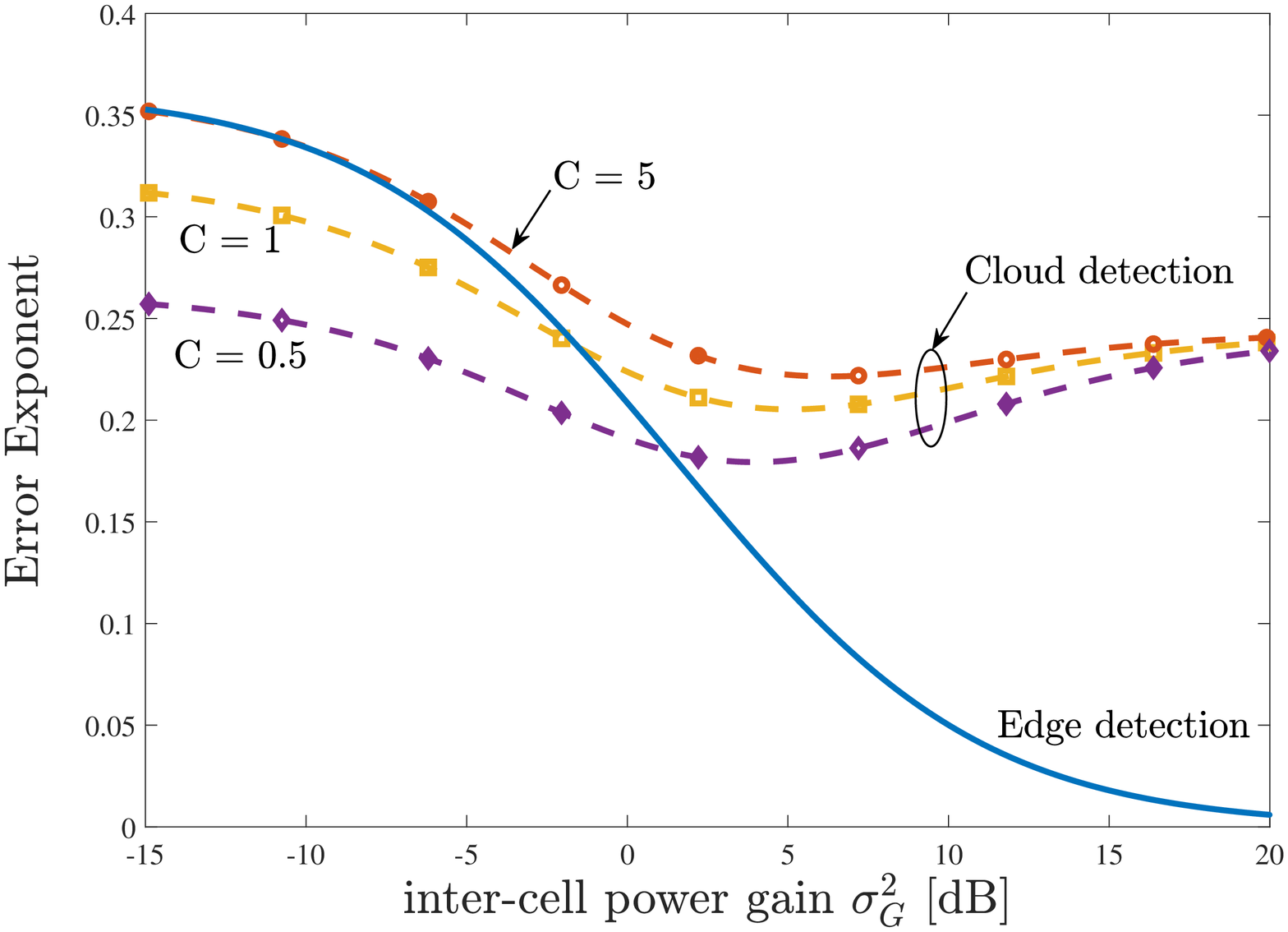}
	\caption{Error exponent for edge and cloud detection as function of the inter-cell power gain $\sigma^2_G$ ($\mu_H=1,\sigma^2_H=1,\mu_G=0$, $\lambda=4$, and $\mathrm{SNR}=3\ \mathrm{dB}$).}
	\label{fig:expo_sigma2G}
\end{figure}
\par \textbf{Asymptotic analysis: }In Fig.~\ref{fig:expo_sigma2G}, we plot the error exponent derived in Sec. \ref{sec:asymptotic_performance} for both edge and cloud detection as a function of the inter-cell power gain $\sigma^2_G$. The performance of edge detection is seen to decrease, i.e., the error exponent decreases, when the inter-cell gain increases. This is due to the fact that the QoI in the other cell may be different, with non-zero probability, from the QoI in the given cell. When this happens, signals sent from devices in the other cell create interference at the EN in the given cell. In contrast, the performance of cloud detection depends on the inter-cell power gain in a more complex fashion that is akin to the behavior of the sum rate in cellular systems with cloud-based decoding \cite{simeone2012cooperative}. In fact, joint detection at the cloud treats as useful the signal received by both cells. Therefore, as long as the inter-cell interference power is large enough, having an additional signal path to the cloud through the other EN can improve the detection performance. This is not the case for smaller values of $\sigma^2_G$, in which case the potentially deleterious effect of inter-cell interference is not compensated by the benefits accrued via joint decoding on the detection of the QoI of the other cell.\par
In Fig.~\ref{fig:expo_sigma2G}, the performance of cloud detection is also seen to depend strongly on the values of the fronthaul capacity $C$. When $C$ is small enough, making fronthaul quantization noise significant, cloud detection can in fact be outperformed by edge detection. In contrast, if $C$ is sufficiently large, edge and cloud detection have the same performance when $\sigma^2_G$ is small, in which case no benefits can be accrued via joint decoding at the cloud, but cloud detection can vastly outperform edge detection when $\sigma^2_G$ is large enough.
\begin{figure}[h]
	\centering
	\includegraphics[height= 6.3 cm, width= 9.5 cm]{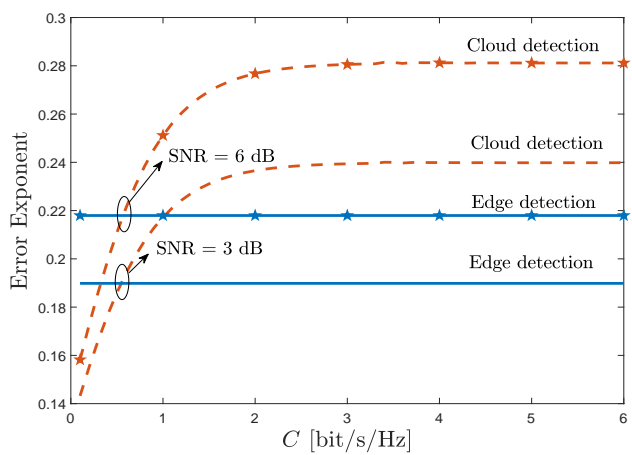}
	\caption{Error exponent for edge and cloud detection as function of the fronthaul capacity $C$ ($\mu_H=1,\sigma^2_H=1,\mu_G=0, \sigma^2_G=1$, and $\lambda=4$).}
	\label{fig:expo_C}
\end{figure}
\par The role of the fronthaul capacity in determining the relative performance of the edge and cloud detection is further explored in Fig.~\ref{fig:expo_C}, where we plot the error exponent as function of the fronthaul capacity $C$ for two different values of the SNR. Consistently with the discussion above, the cloud's detection performance is observed to increase with the fronthaul capacity, outperforming edge detection for large enough $C$. Furthermore, the threshold value of $C$ at which cloud detection outperforms edge detection is as low as $1\ \mathrm{bit/s/Hz}$.
\begin{figure}[h]
	\centering
	\includegraphics[height= 6.3 cm, width= 9.5 cm]{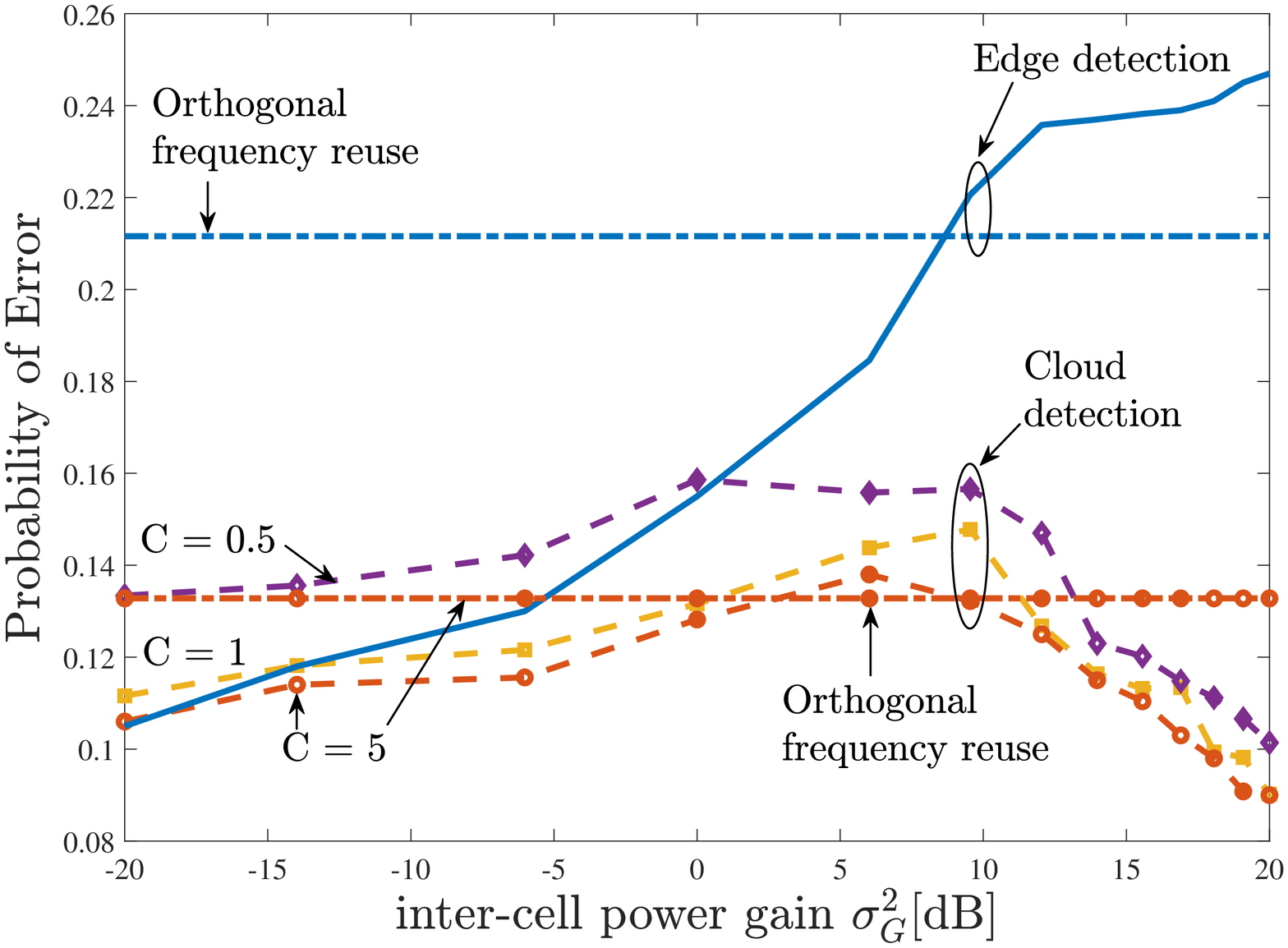}
	\caption{Probability of error for edge and cloud detection as function of $\sigma^2_G$ ($\mu_H=1,\sigma^2_H=1,\mu_G=1, \sigma^2_G=1$, $\mathrm{SNR}=3\ \mathrm{dB}$, $L=5$ and $\lambda=4$).}
	\label{fig:Pe_sigmaG}
\end{figure}
\par 
\textbf{Probability of error for optimal detection: }We now validate the results from the analysis by evaluating the probability of error of the optimal detectors described in Sec. \ref{sec:optimal_detection} via Monte Carlo simulations. Throughout, we set $L=5$ collections. We start in Fig.~\ref{fig:Pe_sigmaG} by plotting the probability of error as a function of the inter-cell power gain $\sigma^2_G$. In a manner consistent with the analytical results illustrated in Fig.~\ref{fig:expo_sigma2G}, the probability of error for edge detection with non-orthogonal frequency reuse is seen to increase when the interference's power increases. In contrast, for cloud detection, the probability of error grows larger with an increasing inter-cell gain for smaller values of $\sigma^2_G$, and then it decreases gradually for higher values of $\sigma^2_G$ as the inter-cell signals become beneficial for joint detection at the cloud.\par
In Fig.~\ref{fig:Pe_sigmaG}, we also compare the performance of non-orthogonal frequency reuse in all cells, which has been assumed thus far, with orthogonal frequency reuse. For edge detection, orthogonal frequency reuse outperforms non-orthogonal frequency reuse for high inter-cell interference power, in which regime the rate gain of having more radio resources in the non-orthogonal reuse scheme is outweighted by the absence of interference with the orthogonal scheme. In contrast, for cloud detection, for high enough inter-cell power, inter-cell signals become useful thanks to joint decoding, and thus, non-orthogonal frequency reuse outperforms orthogonal frequency reuse. 
\begin{figure}[h]
	\centering
	\includegraphics[height= 6.3 cm, width= 9.5 cm]{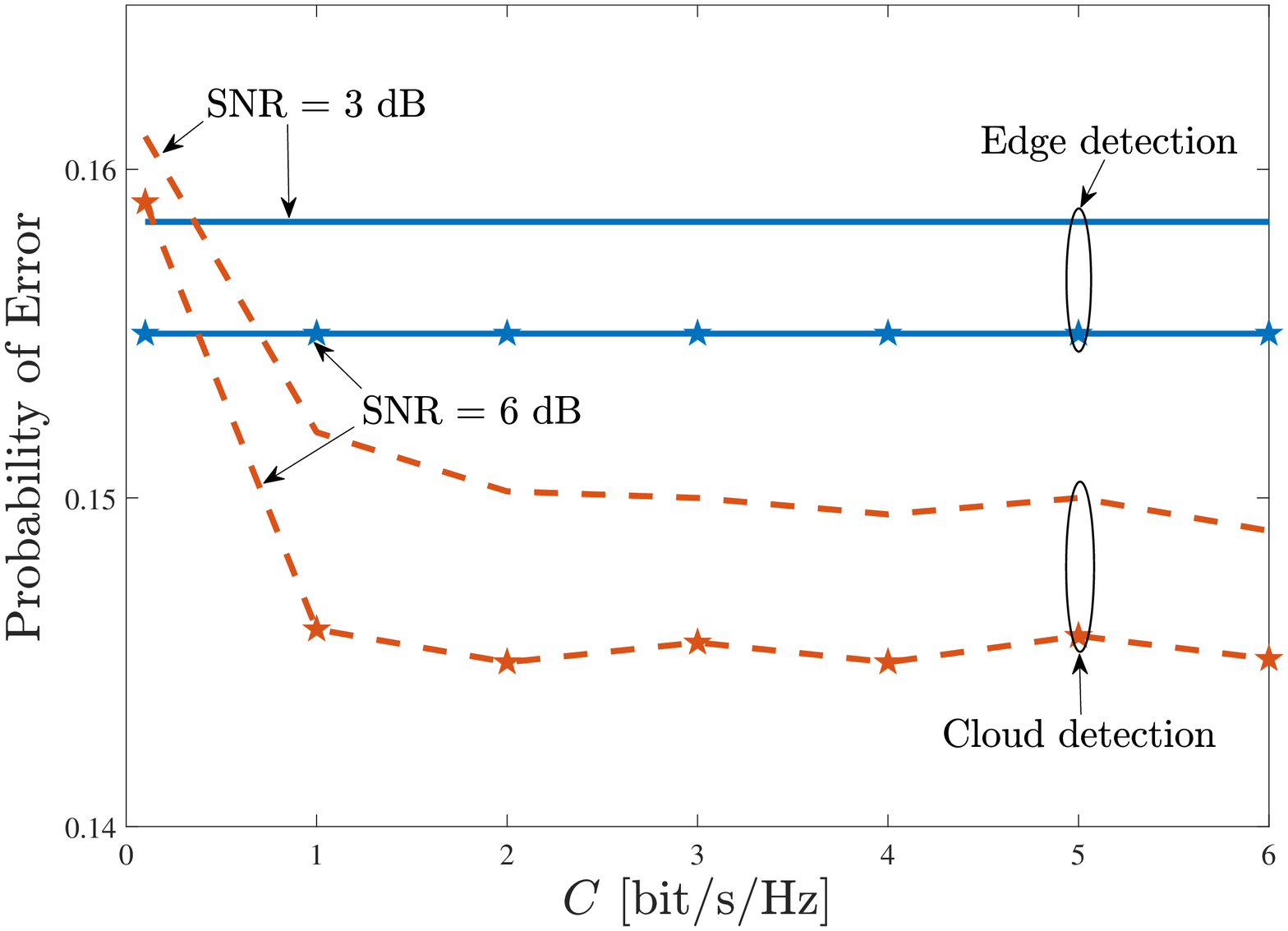}
	\caption{Probability of error for optimal edge and cloud detection as function of $C$ ($\mu_H=1,\sigma^2_H=1,\mu_G=1, \sigma^2_G=1$, $L=5$, $\rho = 0.85$ and $\lambda=4$).}
	\label{fig:Pe_C}
\end{figure}
\begin{figure}[h]
	\centering
	\includegraphics[height= 6.3 cm, width= 9.5 cm]{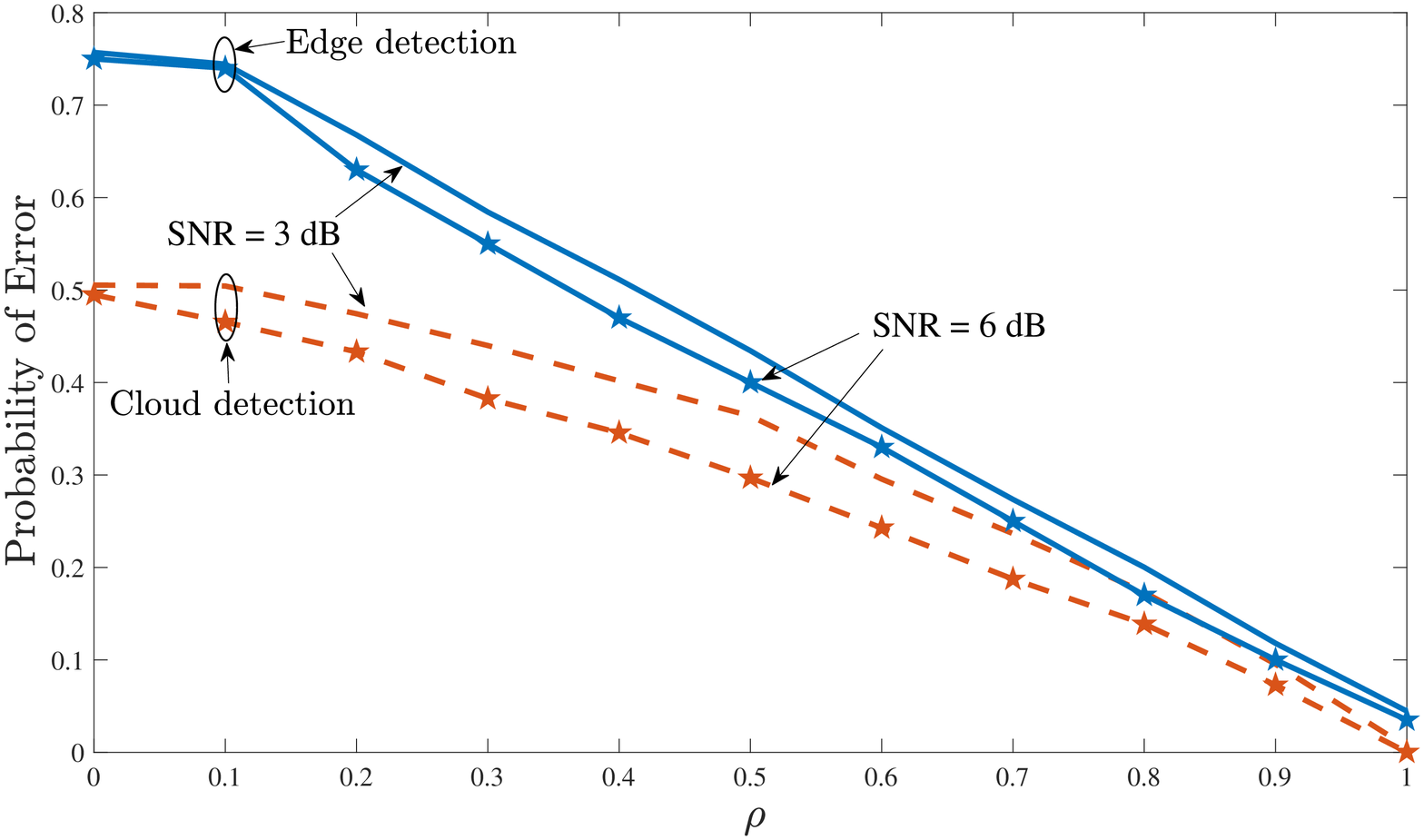}
	\caption{Probability of error for edge and cloud detection using both learning and optimal detection as function of the correlation $\rho$ between the two QoIs in the two cells ($C=5$, $\mu_H=1,\ \sigma^2_H=1,\ \mu_G=1,\ \sigma^2_G=1$, $\mathrm{SNR}=3\ \mathrm{dB}$, $L=5$ and $\lambda=4$).}
	\label{fig:Pe_rho}
\end{figure}

\begin{figure}[h]
	\centering
	\includegraphics[height= 6.3 cm, width= 9.5 cm]{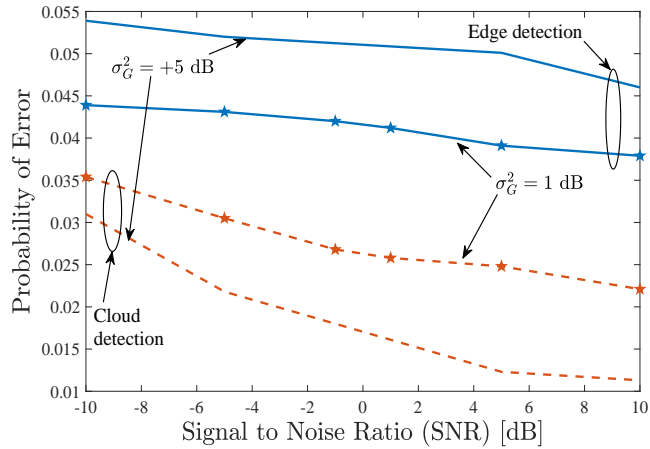}
	\caption{\textcolor{black}{Probability of error for edge and cloud detection using optimal detection as function of the signal to noise ratio (SNR) ($C=10$, $\mu_H=1,\ \sigma^2_H=1,\ \mu_G=1,\ \sigma^2_G=1$, $L=10$ and $\lambda=8$).}}
	\label{fig:Pe_SNR}
\end{figure}
\par We now study the impact of the fronthaul capacity $C$ by plotting the probability of error for optimal edge and cloud detection as function of $C$ in Fig. \ref{fig:Pe_C}. Confirming the discussion based on the asymptotic analysis considered in Fig.~\ref{fig:expo_C}, we observe that the probability of error for optimal cloud detection decreases as function of the fronthaul capacity, and, for a large enough value of $C$, cloud detection is able to outperform edge detection.
\par Since the asymptotic analysis is insensitive to the value of the QoI correlation parameter $\rho$, in Fig.~\ref{fig:Pe_rho}, we evaluate the impact of $\rho$ by studying the probability of error as function of $\rho$ for both optimal edge and cloud detection. For $\rho = 0$, the QoIs in the two cells have opposite values with probability one. Therefore, given the large value of the inter-cell gain, the signals received at the ENs are close to being statistically indistinguishable under the two possible hypotheses $(\theta^1 = \theta_0,\theta^2= \theta_1)$ and $(\theta^1 = \theta_1,\theta^2= \theta_0)$. In contrast, when $\rho$ increases, the two QoIs are more likely to have the same value, decreasing the probability of error for both cloud and edge. \textcolor{black}{Note that, even for $\rho=0.5$, which corresponds to independent QoIs, cloud detection can improve over edge detection. This is because the lack of correlation between the QoIs does not remove the advantage of joint processing of the interfering signals from different cells.}\\
\begin{figure}[t]
	\centering
	\includegraphics[height= 6.3 cm, width= 9.5 cm]{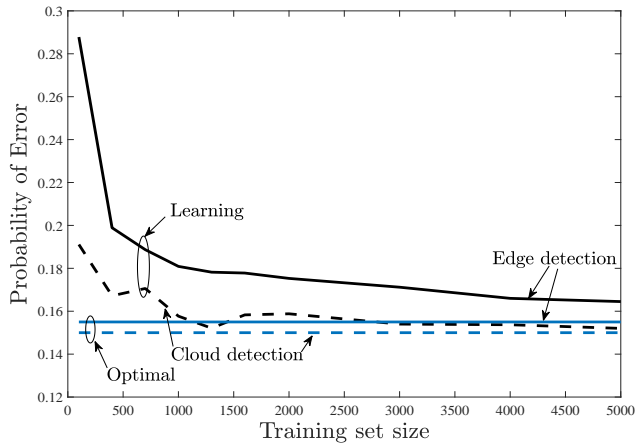}
	\caption{Probability of error for edge and cloud detection using learning as function of the training set size ($C=5$, $\mu_H=1,\sigma^2_H=1,\mu_G=1, \sigma^2_G=1$, $\mathrm{SNR}=3\ \mathrm{dB}$, $\rho=0.85$, and $\lambda=4$).}
	\label{fig:Pe_training_size}
\end{figure}
\textcolor{black}{ In Fig. \ref{fig:Pe_SNR}, we plot the probability of error as function of the SNR for both edge and cloud detection for $L=10$, $\lambda=8$ and two different values of the inter-cell power gain $\sigma^2_G$. Confirming the theoretical conclusions in the paper, we observe that increasing the inter-cell power gain decreases the probability of error for cloud detection, which is not interference limited. In contrast, the performance of edge detection does not improve significantly with larger SNR values, due to the limitations caused by inter-cell interference. }\\

\textbf{Edge and cloud learning: }
We now evaluate the performance of learning-based detection as a function of the size $N$ of the available training set. Training is done using scaled conjugate gradient backpropagation on the cross-entropy loss, as proposed in \cite{moller1993scaled} and implemented in MATLAB's Deep Learning tool box \footnote{https://www.mathworks.com/products/deep-learning.html} with fixed learning rate equal to $0.01$. In Fig.~\ref{fig:Pe_training_size}, we plot the probability of error for both edge and cloud detection using the optimal and learning-based detection techniques as function of $N$. For both edge and cloud detection, the probability of error decreases as function of the training set size until it approximates closely the optimal detector's probability of error. The key observations in Fig.~\ref{fig:Pe_training_size} is that the probability of error for cloud learning converges faster than edge learning to the optimal error. Even though the cloud detector performs a quaternary hypothesis testing problem, its operation in a larger domain space makes it easier to train an effective detector. This is particularly the case for large correlation coefficients, here $\rho = 0.85$, since this implies that two hypotheses, namely, $\mathcal{H}_{00}$ and $\mathcal{H}_{11}$, have a significantly higher prior probability than the remaining two hypotheses.
\section{Conclusions and Extensions}
\label{sec:generalization}
This paper considers the problem of detecting correlated quantities of interest (QoIs) in a multi-cell Fog-Radio Access Network (F-RAN) architecture. An information-centric grant-free access scheme is proposed that combines Type-Based Multiple Access (TBMA) \cite{anandkumar2007type} with inter-cell non-orthogonal frequency reuse scheme. For this scheme, detecting QoIs at the cloud via a fronthaul-aided network architecture was found to be advantageous over separate edge detection for high enough fronthaul capacity in the presence of sufficiently large inter-cell power gains. This is because cloud detection can benefit from inter-cell interference via joint decoding when the correlation between QoIs among different cells is high enough thanks to TBMA. The latter observation was also verified analytically for the asymptotic regime when the number of measurement collections from devices goes to infinity. Under the same conditions, cloud detection was seen via numerical results to outperform edge detection even without model information in the presence of limited data used for supervised learning.\par
Finally, the proposed protocol can be implemented by using the random access preambles from the standard cellular protocols. 
Hence, this form of TBMA changes only the interpretation of those preambles, which means that it can be implemented without intervention on the physical layer of the existing IoT devices.\par
Some extensions and open problems are discussed next. 
First, it would be interesting to consider QoIs with more than two values and multi-cell network with more than two cells. The analysis of this scenario follows directly from the derivations in this paper at the cost of a significantly more cumbersome notation.
\textcolor{black}{
To briefly elaborate on this point, assume that each QoI in each cell $c$ can take $Q$ values, i.e., $\theta^c \in \{\theta^c_1,\ldots, \theta^c_Q\}$ and that there are $K$ cells. In this case, each EN performs a $Q$-ary hypothesis test to distinguish among the $Q$ hypotheses $\mathcal{H}^c_q: \theta^c = \theta^c_q$ for $q \in  \{1,\ldots, Q \}$. The optimal test for deciding among multiple hypothesis at the edge is given by the MAP rule 
\begin{equation}
            \argmax_{q \in \{1, \ldots, Q\} }\Bigg\{ \log p(\theta_q) +  \sum_{l=1}^{L} \log f(\mathbf{Y}^c_{l} | \theta^c=\theta_q) \Bigg\}, \label{eq:optimal_edge_multiple_QoI}
\end{equation} 
which generalizes \eqref{eq:optimal_edge}. The optimal cloud detector aims to solve the $Q^K$-ary hypothesis testing problem among hypotheses $\mathcal{H}_{q_1,\ldots, q_K}:(\theta^1, \ldots, \theta^K)=(\theta^1_{q_1},\ldots,\theta^K_{q_K})$ for $q_k \in \{1,\ldots,Q\}$. The optimal detector in this case can be written as the MAP rule
\begin{equation}
\begin{aligned}    
     \argmax_{(q_1,\ldots, q_K) \in \{1, \ldots, Q\}^{K} } & \Bigg\{ \log  p(\theta_{q_1},\ldots,\theta_{q_K}) \\ & +  \sum_{l=1}^{L} \log f(\hat{\mathbf{Y}}_{l} | \theta^1=\theta_{q_1}, \ldots, \theta^K=\theta_{q_K}) \Bigg\}, \label{eq:optimal_cloud_multiple_QoI}
\end{aligned}    
\end{equation}
which generalizes \eqref{eq:optimal_cloud}. Analysis of \eqref{eq:optimal_edge_multiple_QoI}-\eqref{eq:optimal_cloud_multiple_QoI} can now be carried out by following the same steps in the paper via Chernoff information and the union bound.}
\par \textcolor{black}{Second, an interesting extension would be to study} the design of optimized quantizers between analog observations and discrete levels used for grant-free access.\par
\textcolor{black}{Third,} another interesting direction of research, following \cite{popovski2018slicing}\cite{rahif_access_2018}, is to consider the coexistence of IoT devices with other 5G services, most notably eMBB and URLLC. While orthogonal resource allocation among services would yield separate design problems, non-orthogonal multiple access across different services was found to be advantageous in \cite{popovski2018slicing} \cite{rahif_access_2018}. As a brief note on this problem, in contrast with the sporadic and short IoT transmissions, eMBB transmissions typically span multiple time slots \cite{5gtutorial}. Accordingly, from each IoT device point of view, eMBB signals may be treated as an additional source of noise. However, IoT signals may be decoded and cancelled prior to eMBB decoding \cite{popovski2018slicing}. Like IoT traffic, URLLC traffic is instead typically sporadic and hard to predict. Detectors should hence be designed in order to adapt to the possible presence of URLLC signals. As for URLLC transmissions, the key issue is guaranteeing high reliability despite interference from IoT signals. \par
\textcolor{black}{Fourth, the comparison of the error exponents of edge and cloud detection was done in terms of lower bounds based on the union bound. A more fundamental investigation would account for the tightness of such bounds.}\par
\textcolor{black}{Finally, it would be interesting to generalize the setup to include different \enquote{types} of QoIs (for e.g., pollution and humidity levels). In this case, TBMA as used in this paper will fall short, as devices measuring different types of QoIs cannot be differentiated. To adapt TBMA for this scenario, one could use different codewords for distinct measurements made by sensors. For instance, devices measuring one QoI (e.g., pollution level) in one part of the cell may use a set of codewords, while devices measuring the other QoI (e.g., humidity level) in another part of the cell may use a different set. Note that these sets may not be orthogonal if the receiver used a more sophisticated decoder, for e.g., based on a Bayesian formulation. A similar approach was recently investigated in \cite{JSC_MP_grant_free_IoT}. }
\appendix
\subsection{Proof of Lemma 1}
\label{sec:appendix_1}
The mutual information term in \eqref{eq:capacity_contraint} can be written as 
\begin{equation}
\begin{aligned}
    I(\mathbf{Y}^c_l ; \mathbf{Y}^c_l + \mathbf{Q}^c_l) & =  h(\mathbf{Y}^c_l + \mathbf{Q}^c_l) - M \log (2 \pi \sigma^2_{q^c}), \label{eq:p1_eq1}
    \end{aligned}
\end{equation}
where the equality follows from the assumption that the quantization noises are Gaussian and independent across all observations.
The first term in equation \eqref{eq:p1_eq1} can be bounded as
\begin{equation}
     h(\mathbf{Y}^c_l + \mathbf{Q}^c_l) \leq  \log(2\pi e |\mathbf{\Sigma}_{\mathbf{Y}^c_l} + \sigma^2_{q^c} \mathbf{I}|),\label{eq:p1_eq2}
\end{equation}
where $\mathbf{\Sigma}_{\mathbf{Y}^c_l}$ is the covariance matrix of vector $\mathbf{Y}^c_l$. The inequality follows by the property of the Gaussian distribution of maximizing the differential entropy under a covariance constraint \cite{cover2012elements}. Using the law of iterated expectations, the covariance $\mathbf{\Sigma}_{\mathbf{Y}^c_l}$ can be written as 
\begin{equation}
    \mathbf{\Sigma}_{\mathbf{Y}^c_l} = \sum_{j=0}^{1} \sum_{k=0}^{1} \mathrm{Pr}(\theta^1 = \theta_j,\theta^2 = \theta_k) \mathbf{\Sigma}^c_{j,k}, 
\end{equation}
where matrices $\mathbf{\Sigma}^c_{j,k}$ are diagonal and represent the covariance matrices of $\mathbf{Y}^c_l$ when hypothesis $\theta^c = \theta_j$ and $\theta^{c^\prime} = \theta_k$ hold as defined in \textit{Proposition 1}. This concludes the proof.
\qed
\subsection{Proof of Proposition 1}
\label{sec:appendix_a}
From the union bound $P_e \leq P_e^1 + P_e^2$ with $P_e^c = \mathrm{Pr}[\hat{\theta}^c \neq \theta^c]$ and the identity $P_e^c = \frac{1}{2} \mathrm{Pr}[\hat{\theta}^c \neq \theta^c | \theta^{c^\prime} = \theta_0] + \frac{1}{2}  \mathrm{Pr}[\hat{\theta}^c \neq \theta^c | \theta^{c^\prime} = \theta_1]$, we directly obtain the lower bound on the error exponent
\begin{equation}
     E \geq   \min_{c \in \{0,1 \}} \min_{k \in \{0,1 \}} E^c_k , \label{eq:E_edge_bound}
\end{equation}
where $E^c_k=- \lim_{L \to \infty} \frac{1}{L} \log \mathrm{Pr}[\hat{\theta}^c \neq \theta^c | \theta^{c^\prime}=\theta_k]$ is the error exponent for detection of QoI $\theta^c$ conditioned on the condition $\theta^{c^\prime}=\theta_k$.
Under the optimal Bayesian detector \eqref{eq:optimal_edge}, the detection error exponent $E^c_k$ is given by the Chernoff information \cite[Chapter 11]{cover2012elements} as 
\begin{equation}
    E^c_k = C(f_{0,k}(\mathbf{Y}^c_l),f_{1,k}(\mathbf{Y}^c_l)), \label{eq:E_edge_chernoff}
\end{equation}
where we have denoted $f_{j,k}(\mathbf{Y}^c_l) = f(\mathbf{Y}^c_l|\theta^c = \theta_j , \theta^{c^\prime} = \theta_k)$ for brevity. Computing the error exponent in \eqref{eq:E_edge_chernoff} requires finding the distributions $f_{j,k}(\mathbf{Y}^c_l)$ for $j,k \in \{0,1 \}$. Following \cite{anandkumar2007type}, this can be approximated by a Gaussian distribution in the regime of large $\lambda$ thanks to the Central Limit Theorem (CLT) with random number of summands \cite[p. 369]{billingsley2008probability}.
In particular, referring to \cite{anandkumar2007type} for details, we can conclude that, when $\lambda \to \infty$, the conditional distribution $f_{j,k}(\mathbf{Y}^c)$ tends in distribution to $\mathcal{CN}(\boldsymbol{\mu}_{j,k} , \mathbf{\Sigma}_{j,k})$, where $\boldsymbol{\mu}_{j,k}$ and $\mathbf{\Sigma}_{j,k}$ are the mean vector and covariance matrix respectively when $\theta^c = \theta_j$ and $\theta^{c^\prime}=\theta_k$ and are defined in \eqref{eq:dist_edge} and \eqref{eq:BigSigma}.
\par The Chernoff Information between two Gaussian distributions can be obtained by maximizing over $\alpha \in [0,1]$ the $\alpha$-Chernoff information defined as \cite{nielsen2011chernoff}
\begin{equation}
\begin{aligned}
    & C_{\alpha}(f_{0,k}({\mathbf{Y}}^c_l),f_{1,k}({\mathbf{Y}}^c_l)) =\\ &\frac{1}{2} \mathrm{log} \frac{|\alpha \mathbf{\Sigma}_{0,k} + (1-\alpha)\mathbf{\Sigma}_{1,k}|}{|\mathbf{\Sigma}_{0,k}|^\alpha|\mathbf{\Sigma}_{1,k}|^{1-\alpha}} + \frac{\alpha (1-\alpha)}{2}(\boldsymbol{\mu}_{0,k} - \boldsymbol{\mu}_{1,k})^{\mathsf{T}}\\ & \times (\alpha \mathbf{\Sigma}_{0,k} + (1-\alpha)\mathbf{\Sigma}_{1,k})^{-1} (\boldsymbol{\mu}_{0,k} - \boldsymbol{\mu}_{1,k}). \label{eq:alpha_chernoff}
\end{aligned}
\end{equation} 
By plugging in \eqref{eq:E_edge_bound} and \eqref{eq:alpha_chernoff} the expressions of $\boldsymbol{\mu}_{j,k}$ and $\mathbf{\Sigma}_{j,k}$ and using \eqref{eq:E_edge_chernoff} we obtain the desired result.\qed
\subsection{Proof of Proposition 2}
\label{sec:appendix_b}
Using the law of iterated expectation, the error probability can be written as 
\begin{equation}
    \mathrm{P}_e = \!\!\! \sum_{j,k \in \{ 0,1\}} P(\theta^c = \theta_j, \theta^{c^\prime}=\theta_k) \mathrm{P}_{e | \mathcal{H}_{jk}}, \label{eq:P_e_2}
\end{equation}
where 
\begin{equation}
    \mathrm{P}_{e | \mathcal{H}_{jk}} =\!\!\! \sum_{\{j^\prime,k^\prime\} \neq \{j,k\}} \!\!\! \mathrm{Pr}(\hat{\theta}^c = \theta_{j^\prime},\hat{\theta}^{c^\prime} = \theta_{k^\prime}|\theta^c = \theta_j,\theta^c = \theta_k) \label{eq:P_e_j}
\end{equation}
is the probability of error when hypothesis $\mathcal{H}_{j,k}$ holds, i.e., $\theta^c = \theta_j$ and $\theta^{c^\prime} = \theta_k$. Furthermore, defining the log-likelihood 
\begin{equation}
    L_{jk}(\hat{\mathbf{Y}}_l) \!\!\!= \max_{j,k \in \{0,1 \}} \Big[ \log f_{j,k}(\Hat{\mathbf{Y}}^c_l)+ \log \mathrm{Pr}(\theta^c = \theta_j, \theta^{c^\prime}=\theta_k) \Big],
\end{equation}
we have 
\begin{equation}
\begin{aligned}
        & \mathrm{Pr}[\hat{\theta}^c = \theta_{j^\prime},\hat{\theta}^{c^\prime}= \theta_{k^\prime} \stretchrel{\mid}{  \theta^c = \theta_j , \theta^{c^\prime}=\theta_k]}\\ &= \mathrm{Pr} \Big[ L_{j^\prime k^\prime} (\hat{\mathbf{Y}}_l) \geq \max_{ \{j^{\prime \prime}k^{\prime \prime} \} \neq \{j^\prime k^\prime\}} L_{j^{\prime \prime} k^{\prime \prime}} (\hat{\mathbf{Y}}_l) \stretchrel{\mid}{  \theta^c \!\!\!= \theta_j , \theta^{c^\prime}\!\!\!=\theta_k }\Big]\\
        & \leq \mathrm{Pr} \Big[ L_{j^\prime k^\prime}(\hat{\mathbf{Y}}^c_l) \geq  L_{jk}(\hat{\mathbf{Y}}^c_l) | \theta^c = \theta_j , \theta^{c^\prime}=\theta_k \Big]\\
        & =  \mathrm{Pr} \Big[ \log \frac{f_{{j^\prime k^\prime}}(\hat{\mathbf{Y}}^c_l)}{f_{{jk}}(\hat{\mathbf{Y}}^c_l)}\geq \log \frac{\mathrm{Pr}(\theta^c = \theta_j , \theta^{c^\prime}=\theta_k)}{\mathrm{Pr}(\theta^c = \theta_{j^\prime} , \theta^{c^\prime}=\theta_{k^\prime})} \Big]\\
        & =  e^{-L D(f_{{j^\prime k^\prime}}^{\star} || f_{{jk}}) + \mathcal{O}(L)}, \label{eq:P_jjprime}
\end{aligned}
\end{equation}
where the last equality follows from Sanov's Theorem \cite[p. 362]{cover2012elements} with $f_{{j^\prime k^\prime}}^{\star}(\mathbf{Y}) \propto f_{{j^\prime k^\prime}}^{\lambda} (\mathbf{Y}) f_{{jk}}^{1-
\lambda}(\mathbf{Y})$ and $\lambda$ chosen to satisfy the equality  
\begin{equation}
  D(f^{\star} || f_{{jk}}) - D(f^{\star} || f_{{j^\prime k^\prime}}) = (1/L) \log \frac{\mathrm{Pr}(\theta^c = \theta_j , \theta^{c^\prime}=\theta_k)}{\mathrm{Pr}(\theta^c = \theta_{j^\prime} , \theta^{c^\prime}=\theta_{k^\prime})}. \label{eq:condition_lambda}
\end{equation}
For $L \to \infty$, using \eqref{eq:condition_lambda} and the relation between KL divergences and Chernoff information we obtain \cite{cover2012elements}
\begin{equation}
    D(f^{\star} || f_{{jk}})= D(f^{\star} || f_{{j^\prime k^\prime}}) = C(f_{{j^\prime k^\prime}} || f_{{jk}}) = C(f_{{jk}} || f_{{j^\prime k^\prime}}). \label{eq:chernoff_equality}
\end{equation}
Finally using \eqref{eq:P_e_j}, \eqref{eq:P_jjprime} and \eqref{eq:chernoff_equality}, the probability of error \eqref{eq:P_e_2} can be bounded as
\begin{equation}
    P_e \leq \sum_{j,k \in \{0,1 \}} \!\!\!\mathrm{Pr}(\theta^c=\theta_j,\theta^{c^\prime}=\theta_k) \!\!\!\sum_{j^{\prime} k^{\prime} \neq j,k} \!\!\!\! e^{-L C(f_{{j^\prime k^\prime}}||f_{{jk}}) + o(L)}. \label{eq:P_e_chernoff}
\end{equation}
The proof is then concluded as for \textit{Proposition 1} by invoking the CLT with random number of summands.\qed
\textcolor{black}{
\subsection{Proof of Theorem 1}
To prove the limit in \eqref{eq:theorem_1}, we show that the limits $ \lim_{\sigma^2_G \to \infty }E^c = 0\ \text{hold for}\ c \in \{1,2\}$. To this end, we observe from \eqref{eq:expo_edge} that the first term in $E^c$ tends to zero since, from \eqref{eq:BigSigma}, its limit equals
\begin{equation}
\lim_{\sigma^2_G \to \infty} \sum_{m=1}^{M} \log \Big(\frac{\sigma^2_G \lambda p^c_k(m)}{(\sigma^2_G \lambda p^c_k(m))^{\alpha + 1 - \alpha}} \Big) = 0.
\end{equation}{}
A similar argument applies to the second term in \eqref{eq:expo_edge}, whose limit equals
\begin{equation}
    \lim_{\sigma^2_G \to \infty} \frac{\alpha(1- \alpha)}{2} \sum_{m=1}^{M} \frac{(\mu_{1,k}^c(m) - \mu_{2,k}^c (m))^2}{\sigma^2_G \lambda p^c_k(m)} = 0.
\end{equation}{}}\par
\textcolor{black}{Moving to the cloud's error exponent $E^{cloud}$, we start by characterizing the asymptotic behaviour of the quantization noise when $\sigma^2_G \to \infty$.}\par 
\textcolor{black}{\textit{Lemma 2}: The fronthaul quantization noise for any cell $c\in \{1,2 \}$ satisfies the following limit
\begin{equation}
\lim_{\sigma^2_G \to \infty} \frac{\sigma^2_{q^c}}{\sigma^2_G} = \lambda^{\frac{1}{2M^2}}. \label{eq:limit_sigma2G}
\end{equation}
}
\textcolor{black}{\textit{Proof: } Using \textit{Lemma $1$}, we have the following approximation
\begin{equation}
    \begin{aligned}
           MC & = \frac{1}{2} \sum_{m=1}^{M} \log  \\  & \Bigg( \frac{ \sum_{j=0}^{1} \sum_{k=0}^{1} \mathrm{Pr}( \theta^1 = \theta^1_j,\theta^2 = \theta^2_k) \Sigma^{c}_{j,k} (m,m) + \sigma_{q^c}^2}{(\sigma_{q^c}^2)^M} \Bigg). \label{eq:quantization_limit}\\
           & \approx \frac{1}{2} \sum_{m=1}^{M} \log \Bigg( \frac{\sigma^2_G \lambda (p^c_0 (m) + p^c_1(m))}{(\sigma^2_{q^c})^M} \Bigg),
    \end{aligned}
\end{equation}
from which we can directly derive \eqref{eq:limit_sigma2G}.\qed}\par
\textcolor{black}{Using \textit{Lemma 2}, the diagonal elements of each covariance matrix $\mathbf{\Sigma}_{j,k}$ in \eqref{eq:param_cloud_2} satisfy the limits
\begin{equation}
\begin{aligned}
    & \Sigma_{j,k}(m,m)/ \sigma^2_G \to  \lambda p^1_k(m) +  \lambda^{\frac{1}{2M^2}},\mathrm{for}\ m=1,\ldots,M \\
    & \Sigma_{j,k}(m,m)/\sigma^2_G \to \lambda p^2_j(m) + \lambda^{\frac{1}{2M^2}},\mathrm{for}\ m=\!M\!+\!1\!,\!\ldots\!,\!2M\!, \\
    \end{aligned} \label{eq:limit_sigma_th}
\end{equation}
while the off-diagonal elements, being independent of $\sigma^2_G$, are unaffected by the limit. In order to prove that the limit of $E^{cloud}$ is positive, it is enough to show that the expression being optimized in $E_{j,k}$ as per \eqref{eq:expo_cloud} is strictly larger than $0$ for some $\alpha$ and any $\{j^\prime, k^\prime\} \neq \{j,k \}$. This follows because of the positive semi-definiteness of matrix $(\alpha \mathbf{\Sigma}_{j,k} + (1-\alpha)\mathbf{\Sigma}_{j^\prime ,k^\prime})^{-1}$ and the following argument.}\par
\textcolor{black}{First, by \eqref{eq:limit_sigma_th}, matrices $\mathbf{\Sigma}_{j,k}$ and $\mathbf{\Sigma}_{j^{\prime},k^{\prime}}$ are diagonally dominant matrices when $\sigma^2_G \to \infty$ and hence their determinant tends to the product of their diagonal elements, $\Sigma_{j,k}(m,m)$ and $\Sigma_{j^\prime ,k^\prime} (m,m)$ respectively. More formally, we have
\begin{equation}
\begin{aligned}
     \lim_{\sigma^2_G \to \infty} \frac{1}{2} \mathrm{log} & \frac{|\alpha \mathbf{\Sigma}_{j,k} + (1-\alpha)\mathbf{\Sigma}_{j^\prime ,k^\prime}|}{|\mathbf{\Sigma}_{j,k}|^\alpha|\mathbf{\Sigma}_{j^\prime ,k^\prime}|^{1-\alpha}} \\& = 
      \frac{1}{2} \log \Bigg( \prod_{m=1}^{M} \frac{\alpha p^1_k(m) +  (1-\alpha) p^1_{k^\prime} (m)}{(p^1_k(m))^\alpha (p^1_{k^\prime}(m))^{1-\alpha}} \\& \times  \prod_{m=M+1}^{2M} \frac{\alpha p^2_j(m) + (1-\alpha) p^2_{k^\prime} (m)}{(p^2_j(m))^\alpha (p^2_{j^\prime}(m))^{1-\alpha}} \Bigg).
    \end{aligned} \label{eq:Ecloud_first_element_limit}
\end{equation}
Second, each term in the products in \eqref{eq:Ecloud_first_element_limit} is of the form
\begin{equation}
    \frac{\alpha x + (1-\alpha)y}{x^\alpha y^{1-\alpha}} \label{eq:x_y}
\end{equation}{}
with $x = p_k^1(m) \in [0;1]$ and $y = p_{k^\prime}^1(m) \in [0;1]$. Using the weighted arithmetic mean-geometric mean (AM-GM) inequality \cite[pp. 74–75]{cvetkovski_inequalities}, the expression \eqref{eq:x_y} is larger or equal to one with equality when $x=y$. However, given that $\{j,k\}\neq \{j^\prime,k^\prime\}$, there exist at least one term in the products in \eqref{eq:Ecloud_first_element_limit} that is strictly larger than one. This means that \eqref{eq:Ecloud_first_element_limit} is strictly positive, which concludes the proof. \qed}
\bibliographystyle{IEEEtran}
\bibliography{Biblio}

\end{document}